\documentstyle[prb,aps,epsf]{revtex}

\begin{document}

\title{Quantum Collective Creep: a Quasiclassical Langevin Equation
Approach}

\author{Denis~A.~Gorokhov$^1$, Daniel S.~Fisher$^1$, and
Gianni~Blatter$^2$}

\address{$^{1}$Lyman Laboratory of Physics, Harvard University,
Cambridge, MA 02138, USA}

\address{$^{2}$Theoretische Physik, ETH-H\"onggerberg, CH-8093
Z\"urich, Switzerland}

\maketitle

\begin{abstract}

The dynamics of an elastic medium driven through a random medium
by a small applied force is investigated in the low-temperature
limit where quantum fluctuations dominate. The motion proceeds via
tunneling of segments of the manifold through barriers whose size
grows with decreasing driving force $f$. At zero temperature and
in the limit of small drive, the average velocity has the form
$v\propto\exp[-{\rm const.}/\hbar^{\alpha} f^{\mu}]$. For strongly
dissipative dynamics, there is a wide range of forces where the
dissipation dominates and the velocity--force characteristics
takes the form $v\propto\exp[-S(f)/\hbar]$, with $S(f)\propto 1/
f^{(d+2\zeta)/(2-\zeta)}$ the action for a typical tunneling
event, the force dependence being determined by the roughness
exponent $\zeta$ of the $d$-dimensional manifold. This result
agrees with the one obtained via simple scaling considerations.
Surprisingly, for asymptotically low forces or for the case when
the massive dynamics is dominant, the resulting quantum creep law
is {\it not} of the usual form with a rate proportional to
$\exp[-S(f)/\hbar]$; rather we find $v\propto
\exp\{-[S(f)/\hbar]^2\}$ corresponding to $\alpha=2$ and $\mu=
2(d+2\zeta-1)/(2-\zeta)$, with $\mu/2$ the naive scaling exponent
for massive dynamics. Our analysis is based on the quasi-classical
Langevin approximation with a noise obeying the quantum
fluctuation--dissipation theorem. The many space and time scales
involved in the dynamics are treated via a functional
renormalization group analysis related to that used previously to
treat the classical dynamics of such systems. Various potential
difficulties with these approaches to the multi-scale dynamics ---
both classical and quantum --- are raised and questions about the
validity of the results are discussed.

\end{abstract}

\section{Introduction}

Static and dynamic properties of elastic systems in the presence
of disorder have attracted the attention of physicists for more
than three decades. Vortices in superconductors \cite{Blatter},
charge density waves in solids \cite{Gruner}, domain walls in
magnets \cite{Lemerle}, and geological faults \cite{EQ} are
well-known examples of such systems. Mathematically, some of the
problems that arise in modeling are related to those in Burgers
turbulence \cite{Mezard}, stochastic growth of surfaces
\cite{Krug}, and the stability of matter \cite{Lieb}. The physics
of dirty elastic systems thus impacts on several disciplines. In
this paper we focus on the dynamic properties of driven elastic
manifolds at temperatures low enough for quantum fluctuations to
play an important role, in particular, the phenomenon of quantum
creep.

\begin{figure}
\centerline{\epsfxsize = 8.0cm \epsfbox{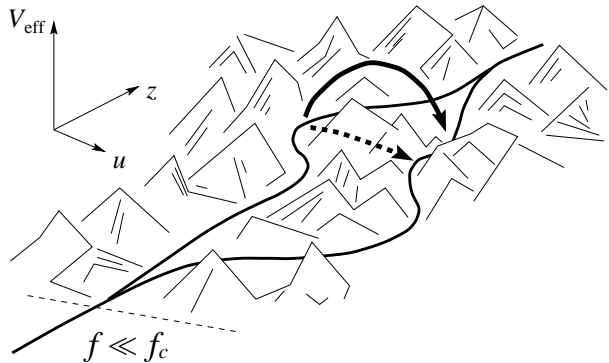}}
\vspace{0.3truecm} \caption{Tunneling (lower dotted line) and activated
motion (upper solid line) of an elastic string in a disordered potential
under the action of a weak external force $f \ll f_c$.}
\label{string}
\end{figure}

In many situations the elastic system can be driven by applied
forces. For example, transport currents in superconductors cause a
Lorentz force to act on vortices, while  a magnetic field applied
to a ferromagnet produces an effective force on domain walls. One
of the primary quantities of interest in such systems is the
average velocity $v$ of the manifold as a function of the applied
force $f$, its dependence on the temperature, and on the magnitude
of the random disorder that impedes the manifold's motion. The
presence of the random pinning forces renders the theoretical
analysis difficult, as perturbation theory breaks down in most of
the important regimes due to the deformations of the manifold on
many length- and time scales. However, some important general
results are known: in the absence of thermal and quantum
fluctuations the system stays pinned --- i.e., the steady state
velocity $v=0$ --- if the driving force $f$ is smaller than a
certain critical value $f_{c}$. At nonzero temperatures, even if
the force $f$ is smaller than $f_{c}$, the system will move with a
nonzero velocity due to thermal creep, see Fig.~\ref{string}. In
the limit $f \ll f_{c}$ the scaling theory of creep
\cite{Feigelman,MPAF,FFH} predicts the law
\begin{equation}
   v \propto \exp\left[-U(f)/T\right],
   \label{classical_law}
\end{equation}
with $U(f)\rightarrow \infty$ as $f\rightarrow 0$ and there is no
linear response to an applied force. The quantity $U(f)$ can be
interpreted as a barrier separating two neighboring minima of the
free energy on some appropriate length scale that depends on the
applied force; on long length- and time scales, the manifold's
motion proceeds via thermally activated jumps of corresponding
segments between different metastable configurations.

Several important assumptions are made in order to derive
Eq.~(\ref{classical_law}). First of all, at least in its simplest
form, it is assumed that the {\it barriers} between different
metastable configurations scale with length in the same way as the
{\it variations} of the free energy that are caused by the balance
between the random pinning and elasticity; in general one could
consider two different scaling laws for these quantities
\cite{FFH}. Second, Eq.~(\ref{classical_law}) is an Arrhenius law
with {\it typical} events dominating the creep-like motion. In
general, {\it rare} events, in particular regions with anomalously
high barriers, might lead to a modification of this law, perhaps
to a form $v \propto \exp\left[-\left(U(f)/T\right)^{\alpha}
\right]$ with $\alpha \ne 1$ an exponent characterizing the tails
of the barrier probability distribution. Because of these
assumptions, as well as the need for increased analytical
understanding, it would be valuable to derive the classical creep
law (\ref{classical_law}) starting from a microscopic description.
Progress in this direction has recently been made, see Refs.\
\onlinecite{Radzihovsky,Chauve}, where Eq.\ (\ref{classical_law})
has been derived from an at-least-partially controllable
renormalization group (RG) expansion. Although these results
support the validity of the scaling theory for thermal creep,
serious questions remain which we will discuss later in this
paper. We also note that neither the recent renormalization group
calculations nor the present paper address an important aspect of
creep in many contexts, in particular in the vortex lattice, i.e.,
the role of dislocations --- the results are, so far, restricted
to a truly elastic manifold with pinning weak enough that
deformations are too small to induce dislocations or other
strongly nonlinear effects \cite{copp-dis}.

Experimental investigations have shown that even at very low
temperatures creep of driven elastic systems, e.g., in vortex
lattices \cite{Yeshurun}, of domain walls in magnets
\cite{domain_wall_experiments}, and of charge-density waves in
solids \cite{ZZ}, still persists. This phenomenon suggests an
explanation in terms of {\it quantum tunneling} of the manifold
between different metastable configurations; the macroscopic
manifestation of this is quantum creep \cite{BGV}. By analogy with
thermal creep, one would guess that
\begin{equation}
   v \propto \exp\left[-S(f)/\hbar\right],
   \label{quantum_law}
\end{equation}
with $S(f)$ a characteristic action for tunneling of a typical
segment of the manifold whose size is determined by the applied
force. In spite of a substantial number of studies of quantum
creep \cite{Yeshurun} there is as yet no theoretical analysis that
starts from a microscopic description of a driven elastic manifold
interacting with impurities; the main goal of this paper is to
present a first start at such an analysis, along with a discussion
of the difficulties involved. We will show that a law of the form
\begin{equation}
   v \propto \exp\left\{-\left[S(f)/\hbar\right]^{\alpha}\right\}
\end{equation}
can be obtained using a renormalization group approach, as can the
functional dependence of the action $S$ on the external force $f$
be found. Whether this is indeed the correct behavior is addressed
at the end of the paper.

A natural tool to investigate quantum creep is the functional
renormalization group (FRG) expansion \cite{Fisher}, an
$\epsilon$-expansion near the upper critical dimension ($d_c=4$)
of the random manifold problem. This involves, intrinsically, an
infinite number of marginal operators that can be combined into
one or more functions. We will find that in this approach a
natural separation of the frequency scales occurs between {\it
inter-valley} and {\it intra-valley} fluctuations of the manifold.
The inter-valley fluctuations correspond to the motion of the
manifold on large scales on which there are many separate valleys
in the ``landscape" caused by the randomness. The intra-valley
fluctuations correspond to the much faster motion of the manifold
within one valley.

The paper is organized as follows. In section~\ref{model} we
formulate the model and find the appropriate effective action. In
section \ref{scaling_arguments} we  show how to derive the creep
law (\ref{quantum_law}) using scaling arguments. Next, we study
the problem using the renormalization group expansion
(section~\ref{renormalization_group_analysis}) and then summarize
and discuss the main results. We end with an analysis of the
limitations and problematic aspects of both our and previous
approaches to the creep problem in section \ref{validity}.

\section{Model and Effective Action}
\label{model}

We describe the elastic medium by a non-dispersive elastic
manifold interacting with a quenched random potential; this
generic model captures all the essential physics while allowing
for an extensive analytical treatment. The {\it classical} motion
of the displacements $u({\bf z},t)$ of the manifold away from its
undeformed state is described by the dynamical equation
\begin{equation}
   \eta \partial_t u
   +\rho \partial^2_t u
   =c{\bf\nabla}^2 u + F(u,{\bf z}) + f + f_{\rm th}({\bf z},t),
   \label{classical_equation}
\end{equation}
where the friction force $\eta\partial_{t}u$ and the inertia $\rho
\partial_t^2 u$ are balanced by the elastic- ($c\Delta u$), random
pinning- ($F(u,{\bf z})$), driving- ($f$), and thermal- ($f_{\rm
th}({\bf z},t)$) forces. We will consider the case, applicable to
domain walls, of $d$-dimensional manifolds in $d+1$-dimensional
random environments, so that ${\bf z}\in {\cal R}^{d}$ and $u \in
{\cal R}^1$.

The random pinning force $F(u,{\bf z})$ is taken to be Gaussian
random with mean zero and a correlator
\begin{equation}
  \overline{F(u,{\bf z}) F(u^{\prime},{\bf z}^{\prime})}
  = \Delta (u-u^{\prime})\, \delta ({\bf z}-{\bf z}^{\prime}),
  \label{FF_correlator}
\end{equation}
where the overbar denotes an average over the disorder. The
function $\Delta (u)$ decays rapidly with a characteristic scale
$\xi$. The thermal noise $f_{\rm th} ({\bf z},t)$ is Gaussian
white with a correlator
\begin{equation}
   \left \langle {f_{\rm th}({\bf z}, t)
   f_{\rm th}({\bf z}^{\prime}, t^{\prime})}\right\rangle
   = 2\eta T \,\delta(t-t^{\prime})
   \delta({\bf z}-{\bf z}^{\prime})
   \equiv \kappa (t-t ^{\prime})\,
   \delta({\bf z}-{\bf z}^{\prime}),
   \label{ff_correlator}
\end{equation}
with $\langle\dots\rangle$ denoting the average over thermal
fluctuations.

In classical statistical mechanics we can reformulate the problem
posed by a stochastic differential equation as an effective field
theory \cite{MSR,Janssen} with the help of the Martin-Siggia-Rose
(MSR) approach. After averaging over thermal fluctuations the
MSR-action corresponding to (\ref{classical_equation}) has the
form
\begin{eqnarray}
   A_{\rm MSR} &=& -\int\thinspace d^{d}z dt \thinspace
   iy({\bf z},t) \big(\eta \partial_t u
   + \rho \partial^2_t u
   - c \Delta {u}\big)\nonumber\\
   &+&\int\thinspace d^{d}z dt \thinspace
   iy \big(f + F(u,{\bf z})\big)\nonumber\\
   &+&\frac{1}{2}\int\thinspace d^{d}zdtdt^{\prime}
   \thinspace iy({\bf z}, t) \kappa (t-t^{\prime})
   iy({\bf z}, t^{\prime}),
   \label{MSR_action}
\end{eqnarray}
with $y({\bf z},t)$ an auxiliary field used to enforce the
equation of motion. The probability of a particular dynamical
evolution $\bar{u}({\bf z},t)$ under the stochastic process is
proportional to $\int{\cal D}[y] \thinspace\exp\{A_{\rm MSR}
[\bar{u},y]\}$.

The main goal of this paper is to investigate the influence of
quantum fluctuations on the system whose {\it classical limit} is
described by Eq.~(\ref{classical_equation}). Again, it is
convenient to formulate the problem as a field-theory, i.e., to
write the effective action describing the elastic system in the
presence of quantum fluctuations in a way analogous to
(\ref{MSR_action}). When calculating time-independent quantities
in equilibrium systems (i.e., in the case $f=0$) the corresponding
quantum action has a Euclidean form and the quantum partition
function can be written as an imaginary time path integral
\cite{FH}. However, here we are interested in non-equilibrium
properties ($f\ne 0$) and the Euclidean action cannot be used.

A formalism allowing to study the {\it real time} dissipative
quantum mechanics of a system is that of Feynman and Vernon
\cite{Feynman}: the quantum amplitude $\Psi (x_f,t_f;x_i,t_i)$ for
a system to have a coordinate $x_f$ at time $t_f$, if at time
$t_i$ it had a coordinate $x_i$, can be written as a path integral
$\int_{x_i, t_i}^{x_f, t_f} {\cal D}[x] \exp(iS[x]/\hbar)$, with
$S[x]$ the classical action. Consequently, the probability of the
transition $x_i,t_i\rightarrow x_f,t_f$ can be written in the form
\begin{equation}
   {\cal P}(x_f, t_f;x_i, t_i) =
   \int_{x_i, t_i}^{x_f, t_f}
   \int_{x_i, t_i}^{x_f, t_f}{\cal D }[x]
   {\cal D} [x^{\prime}]
   \exp(iS[x]/\hbar)
   \exp(-iS[x^{\prime}]/\hbar).
   \label{quantum_partition_function}
\end{equation}
In non-equilibrium quantum mechanics the functional
(\ref{quantum_partition_function}) plays the same role as the
partition function in equilibrium problems. We see that the
effective action can be written as $iS[x]-iS[x^{\prime}]$ and
includes two different paths $[x(t)]$ and $[x^{\prime}(t')]$.

The standard way to include dissipation is to write the action
$S[x]$ in the form $S[x]=S_{0}[u]+S_{\rm bath}[X_{\rm bath}]+
S_{\rm int}[x]$ representing the actions corresponding to the
elastic manifold, the bath, and the interaction between the bath
and the manifold, respectively (note that $x = (u, X_{\rm bath})$
and $x^{\prime} = (u^{\prime}, X^{\prime}_{\rm bath})$ are the
coordinates describing both the manifold and the bath). In the
Caldeira-Leggett model \cite{Caldeira} the terms $S_{\rm
bath}[X_{\rm bath}]$ and $S_{\rm int}[x]$ are taken to be
quadratic in the bath coordinate $X_{\rm bath}$ and hence can be
integrated out at the expense of effective interactions that
couple different times. The resulting action takes the form (see
appendix~\ref{appendix} or Ref.~\onlinecite{Schon})
\begin{eqnarray}
   \frac{iS[{\tilde u},{\tilde y}]}{\hbar}
   &=& -\frac{i}{\hbar}\int\thinspace d^{d}z dt \thinspace
   {\tilde y}({\bf z},t) \left(\eta \partial_t {\tilde u}
   +\rho \partial^2_t {\tilde u}
   -c \Delta {\tilde u}\right )\nonumber\\
   &+&\frac{1}{2\hbar^{2}}\int\thinspace d^{d}zdtdt^{\prime}
   \thinspace
   {i\tilde y}({\bf z}, t)\kappa (t-t^{\prime})
   {i\tilde y}({\bf z}, t^{\prime})\label{quantum_action}\\
   &-&\frac{i}{\hbar}\int d^{d}z dt\Big[U({\tilde u}+
   \frac{{\tilde y}}{2},{\bf z})-
   U({\tilde u}-\frac{{\tilde y}}{2}, {\bf z})-f{\tilde y}\Big],
   \nonumber
\end{eqnarray}
with
\begin{eqnarray}
   {\tilde u}({\bf z},t)&=&(u({\bf z},t)+u^{\prime}({\bf z},t))/2,
   \nonumber \\
   {\tilde y}({\bf z},t)&=&u({\bf z},t)-u^{\prime}({\bf z},t),
   \nonumber
\end{eqnarray}
and the random potential
\[
   U(u, {\bf z})= -\int^{u}\thinspace du_{1} F(u_{1}, {\bf z}).
\]
The Fourier transform of the effective noise correlator
$\kappa(t-t^{\prime})$ of the bath fluctuations obeys the quantum
fluctuation--dissipation theorem,
\begin{equation}
   \kappa (\omega)=\eta\hbar \omega\coth\frac{\hbar\omega}{2T},
   \label{quantum_noise}
\end{equation}
where $\eta$ is related to the spectral density of the bath, see
appendix~\ref{appendix}.

The action (\ref{quantum_action}) allows one, in principle, to
investigate all the properties of a driven elastic medium in the
presence of quantum fluctuations. The classical limit, the MSR
action $A_{\rm MSR}=iS/\hbar$ can be recovered by expanding the
random potential energy $U$ in the ${\tilde y}$-field and
substituting ${\tilde u}$ by $u$ and ${\tilde y}/\hbar$ by $y$.
Such an expansion is valid at high temperatures and it has been
argued \cite{Schmid} that it also applies to the limit
$\hbar\rightarrow 0$ at $T = 0$. This suggests that the dynamics
of quantum systems in the semiclassical regime, where tunneling
processes through large barriers dominate, can be described by
Eq.\ (\ref{classical_equation}) with the noise obeying the
fluctuation--dissipation theorem (\ref{quantum_noise}); this is
called the quasi-classical Langevin equation \cite{Schmid} (QLE)
approach. In a number of papers \cite{Eckern,Koch} the QLE has
been used to study the non-equilibrium dynamics of quantum
systems. In particular, in Ref.~\onlinecite{Eckern} the quantum
tunneling of an overdamped quantum particle has been studied
using this approximation. Writing the inverse lifetime $\Gamma$ of
a metastable state in the form $\Gamma = Pe^{-S/\hbar}$ with $S$
the tunneling action and $P$ the prefactor, it turns out
\cite{Eckern}, that the QLE gives the correct value of the
tunneling action $S$ up to a multiplicative factor of order unity
(note that for quadratic systems the QLE is actually {\it exact},
as is the expansion $U({\tilde u}+ {{\tilde y}}/{2},{\bf z})-
U({\tilde u}- {{\tilde y}}/{2}, {\bf z})\approx U^{\prime} (\tilde
u , {\bf z})\,{\tilde y}$ for this case).

In this paper we use the QLE to study quantum creep in a
disordered medium. Although we would like to do better, we are
primarily interested in obtaining the correct {\it scaling} laws
and asymptotic forms; the exact coefficients are of less interest.
The action (\ref{MSR_action}) with the quantum noise
(\ref{quantum_noise}) is much simpler to analyze than the full
quantum action (\ref{quantum_action}). We believe that the QLE
should be a useful tool for investigating the dynamics of quantum
systems as it appears to capture the essential physics, while
being --- at least relatively --- tractable analytically. Quantum
fluctuations appear in the QLE approach as an effective random
force with a correlator $\kappa (t)$, see (\ref{quantum_noise}),
acting on the system. The quantum decay of a metastable state then
is equivalent to the thermally activated escape of a particle
driven by colored noise \cite{colored}.

\section{Scaling Arguments}
\label{scaling_arguments}

\subsection{Statics and Thermal Creep}
\label{thermal_creep}

An elastic manifold subject to a random potential behaves as an
assembly of approximately independently pinned segments of a
characteristic size that is determined by the balance between the
pinning and elastic forces. The driven motion of the manifold is
dominated, both classically and quantum mechanically, by
successive jumps of segments of the manifold between subsequent
metastable configurations. In order to understand this behavior
one needs to consider various characteristic scales.

The characteristic size of the collectively pinned segments is of
order \cite{Blatter} $L_c\sim [c^2\xi^2/\Delta (0)]^{1/(4-d)}$,
the Larkin length over which the typical distortion of the
manifold $|u(L_c) - u(0)|$ is of order $\xi$, the scale of the
correlations in the random pinning landscape. The characteristic
energy scale of the deformations on scale $L_c$ is $U_c\sim c\xi^2
L_c^{d-2}$. Beyond $L_c$, the typical displacements of the
manifold grow as $|u(L) - u(0)| \sim \xi {\left (L/L_c\right
)}^\zeta$ with $\zeta$ the wandering exponent, $\zeta=(4-d)/3$ for
the random force case, while $\zeta_{d,1} \approx 0.2083\, (4-d)$
for a short-range correlated random potential; the difference in
energy between metastable configurations deviating on a scale $L$
by a displacement $\delta u(L)$ is controlled by the balance
between the elastic- and pinning energies, both of which have a
magnitude of order $U_c{\left (L/L_c\right )}^\theta$ with the
energy exponent $\theta= d+2\zeta - 2$ and the wandering exponent
$\zeta$ given above.

In the presence of an applied force $f$ it is favorable for a
segment of size $L$ to move into the neighboring metastable valley
if the energy $f L^d \delta u(L) \sim fL^d\xi {\left (L/L_c\right
)}^{\zeta}$ gained due to the presence of the force is larger than
the difference of the elastic- plus pinning energies between the
two configurations; for small forces, this will only occur for
large segments. Equating the two energy scales, we obtain the
minimum characteristic size of a segment that can move to the
lower energy site \cite{Feigelman},
\begin{equation}
  L_f\sim L_c {\left (\frac{f_c}{f}\right)}^{1/(2-\zeta )},
  \label{L_f}
\end{equation}
where $f_c\sim c\xi/L_c^2$ is the critical applied force
needed to move the manifold in the absence of thermal or quantum
fluctuations. The scale $L_f$ is the minimum size of a segment
that can hop in order to lower its energy; the macroscopic motion
then proceeds via jumps of segments of this characteristic size
over distances
\begin{equation}
   u_f \sim \xi {\left (\frac{f_c}{f}\right)}^{\zeta/(2-\zeta)}
   \label{u_f}
\end{equation}
separating two neighboring metastable configurations.

Consider first the classical motion, i.e., thermal creep: The
conventional assumption is that the height of the energy barrier
that must be surmounted for such a motion to occur classically is
determined by the scaling of the {\it static} energy  $U_c{\left
(L/L_c\right)}^\theta$ (we will discuss the validity of this
assumption later in this section). The characteristic height of
the barrier that dominates the motion at a small force $f$ then is
\begin{equation}
   U(f) \sim c u_f^2L_f^{d-2}
   \sim U_c {\left (L_f/L_c \right)}^{d+2\zeta -2}
   \sim U_c {\left (f_c/f \right )}^{(d+2\zeta -2)/(2-\zeta)};
   \label{U_f}
\end{equation}
this yields an average velocity for thermal creep, see
Ref.~\onlinecite{Feigelman},
\begin{equation}
   v \sim \exp[-U(f)/T].
   \label{v_f}
\end{equation}

\subsection{Quantum Creep with Dissipative Dynamics}
\label{dissipative_dynamics}

The quantum tunneling of segments of the manifold can be estimated
in a manner analogous to their classical thermal activation: Let
us assume that the manifold is located in one of the metastable
configurations and subject to a small driving force $f$. One
expects the lifetime of such a state to be proportional to
$\exp\left(S/\hbar\right)$, with $S$ the characteristic action
describing the tunneling through the barrier separating the
metastable minimum from one with a lower energy. We estimate $S$
using the standard theory of the decay of metastable states. We
first assume that the effects of dissipation dominate over those
of inertia; the Euclidean action of the manifold then can be
written in the form
\begin{eqnarray}
   S_{\rm Eucl}[u]
   =\int\limits_{-\infty}^{+\infty}d\tau
   \int d^{d}z\bigg[\frac{c}{2}
   {\bigg(\frac{\partial u}{\partial {\bf z}}\bigg)}^{2}
   +U(u,{\bf z})-fu
   +\frac{\eta}{4\pi}\int\limits_{-\infty}^{+\infty}
   d\tau^{\prime}\frac{{\left (u({\bf z},\tau)
   -u({\bf z},\tau^{\prime})\right)}^{2}}
   {{\left (\tau - \tau^{\prime}\right )}^{2}}\bigg],
   \label{euclidean_action}
\end{eqnarray}
with $\tau$ the imaginary time and $\eta$ the dissipative
coefficient. In order to find $S$ one needs to find the imaginary
time trajectory connecting the initial and final configurations of
the manifold and calculate its action. We emphasize that as our
problem is a non-equilibrium one, the full quantum action
(\ref{quantum_action}) should be used. However, at low driving
forces, the problem we study is a {\it quasi-stationary} one, as
the lifetime of a metastable state is large and the manifold can
be considered to be in local equilibrium. Therefore we can use the
simpler (Euclidean) action (\ref{euclidean_action}) instead of the
full dynamic action (\ref{quantum_action}).
\begin{figure}
\centerline{\epsfxsize = 7.0cm \epsfbox{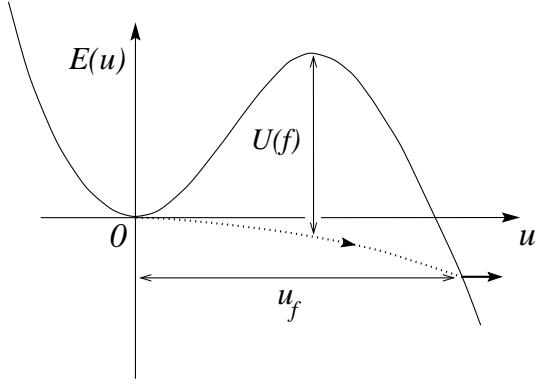}}
\vspace{0.3truecm} \caption{Tunneling of a particle with an
effective mass $\rho L_f^d$ and dissipative coefficient $\eta
L_f^d$ through a barrier of height $U(f)$ and width $L_f$.}
\label{schematic_tunneling}
\end{figure}

In estimating the tunneling action we find the size $L_f$ of the
tunneling segment as in the classical case of thermal creep, since
this is determined by the {\it static} balance between the
variations in the pinning energy and the energy gain due to the
external force. Furthermore, we assume that there is only one
characteristic time scale $\tau_{f}$ associated with the tunneling
process. We can then consider the tunneling event, crudely, as an
effective process in which a point-like object with a friction
coefficient $\eta_{f}=\eta L_{f}^{d}$ tunnels through a potential
barrier of height $U(f)$ and extent $u_{f}$, see
Fig.~\ref{schematic_tunneling}. The saddle-point of the action
(\ref{euclidean_action}) is minimized when
\begin{equation}
   \eta_f \frac{u_f}{\tau_f}\sim \frac{U(f)}{u_f},
\end{equation}
where we have substituted the dynamic term by its characteristic
value and $U(f)/u_{f}$ is the characteristic force acting on the
``particle''. The tunneling time $\tau_{f}$ is given by $\tau_{f}
\sim \eta_{f}u_{f}^{2}/U(f)$, resulting in the tunneling action
\begin{equation}
   S(f) \sim  S_{\eta}
   {\left (\frac{f_{c}}{f}\right )}^{{\left (d+2\zeta\right )}/
   {\left (2-\zeta\right )}},
   \label{dissipative_action_scaling}
\end{equation}
with $S_{\eta}\sim \eta\xi^{2}L_{c}^{d}$ the characteristic scale
of the dissipative action of a segment with the diameter of the
order of the Larkin length $L_c$. The velocity of the manifold is
related to $S(f)$ via (\ref{quantum_law}).

The characteristic crossover temperature $T_{\rm cr}$ from quantum
to classical creep can be obtained by comparing $U(f)/T$ and
$S(f)/\hbar$, i.e.,
\begin{equation}
   T_{cr} \sim \frac{\hbar U(f)}{S(f)}
   \sim \frac{\hbar U_{c}}{S_{\eta}}
   {\left (\frac{f}{f_c}\right)}^{{2}/{(2-\zeta )}}
   \label{crossover_temperature}
\end{equation}
gives the scaling conjecture for the dissipative limit.

\subsection{Quantum Creep with Inertial Dynamics}
\label{inertial_dynamics}

We now assume that the inertial term in the action is more
important for the tunneling process than the dissipation. In this
case the Euclidean action analogous to that given by
(\ref{euclidean_action}) can be written in the form
\begin{equation}
   S_{\rm Eucl}[u]
   =\int\limits_{-\infty}^{+\infty}d\tau\int d^{d}z
   \bigg[\frac{c}{2}
   {\left(\frac{\partial u}{\partial{\bf z}}\right)}^{2}
   +U(u,{\bf z})-fu+\frac{\rho}{2}
   {\left(\frac{\partial u}{\partial t}\right )}^2\bigg],
   \label{euclidean_action_1}
\end{equation}
with $\rho$ the mass density of the manifold. The size $L_f$ of
the tunneling segment is the same as in the dissipative case. In
order to find the action $S(f)$ one again needs to find the
tunneling time $\tau_f$; comparing kinetic and pinning energies
for the saddle-point configuration, we find (with $\rho_f = \rho
L_f^d$ the effective mass of the tunneling object)
\begin{equation}
   \rho_f \frac{u_f}{\tau_f^2}\sim \frac{U(f)}{u_f},
\end{equation}
i.e., $\tau_f\sim {\left [\rho_f/U(f)\right]}^{1/2}u_f$, and the
tunneling action can be estimated as
\begin{equation}
   S(f)\sim S_{\rho}
   {\left (\frac{f_c}{f}\right )}^{(d+2\zeta -1)/(2-\zeta)},
   \label{naive_result}
\end{equation}
with $S_{\rho} \sim (\rho U_c L_c^d)^{1/2}\xi$ the characteristic
action with inertial dynamics on the Larkin scale $L_c$. The
quantities $L_c$ and $U_c$ have been defined in section
\ref{thermal_creep} above. The characteristic crossover
temperature can be determined as in the dissipative case and is
given by $T_{cr}\sim ({\hbar U_c}/{S_{\rho}}){(f/f_c)}^{1/(2-\zeta
)}$.

\subsection{Assumptions}

Two important assumptions have been made in order to derive the
scaling prediction (\ref{quantum_law}): First of all, it has been
assumed that the barriers for the manifold's motion scale in the
same way as the variations of the energy. If there are two
independent exponents $\theta=d-2+2\zeta$ parametrizing the {\it
static} energy scaling and $\psi>\theta$ characterizing the
scaling of energy barriers \cite{activ-scal,SGpsi}, then the
barrier for thermal creep will have the form $U(f)\propto
L_f^{\psi}$. The simplest assumption for the quantum motion is
that this same barrier height is the appropriate one for the
quantum barrier traversal. While the dissipative result
(\ref{dissipative_action_scaling}) would remain the same (as
$S_{\rm diss} \sim \eta_f u_f^2$ does not depend on the height of the tunneling
barrier), the result for the inertial dynamics would take the form
$S_{\rm inertial} \propto L_f^{(d+\psi+2\zeta)/2}$.

We should emphasize, however, that if $\psi>\theta$ then this is
because the dynamics on intermediate length scales affects that on
the large scale $L_f$. A dependence of the large scale dynamics,
of interest for quantum creep at small applied forces, on the
intermediate scale physics can also occur quantum mechanically;
but the way in which it occurs could be quite different from the
classical situation and it is possible that each type of quantum
dynamics has its own exponent $\psi_Q$ distinct from the classical
value $\psi$ \cite{MPAFYoung}. Once we admit the possibility of a
dependence of the tunneling dynamics of large segments on the
dynamics at intermediate scales, the basic assumptions of the
scaling arguments given above break down. But this is exactly the
kind of problem for which a renormalization group framework is
needed.

Another potentially important effect that was not taken into
account in the simple scaling analysis summarized above are rare
events: in general, it is possible that the velocity is not
controlled by the typical barriers --- or, more precisely, by the
tunneling through these typical barriers --- but by rare
anomalously large barriers. In the simplest scenario, this could
give rise to a quantum creep law of the form $v \propto \exp
\left\{-{\left[{S(f)}/{\hbar}\right]}^{\alpha}\right\}$, with
$\alpha \ne 1$ a non-trivial exponent characterizing the tail of
the distribution of barrier heights. Just such a phenomenon
controls the temperature dependence of the diffusion coefficient
of a single particle diffusing in a Gaussian random potential
\cite{Scheidl}, see the discussion in section
\ref{results_for_velocity}. One mechanism for an exponent $\alpha
> 1$ is that rare, rather than typical, events dominate the
macroscopic motion.

\section{Renormalization Group Analysis}
\label{renormalization_group_analysis}

\subsection{Derivation of RG Equations}
\label{derivation_rg_equations}

Having presented the basic scaling arguments for both classical
and quantum creep, we now turn to the main task of this paper, an
attempt to derive the creep laws from a microscopic description of
the dynamics. With the roughness exponent $\zeta$ vanishing in
dimensions $d>4$, one expects that $d_c=4$ is the upper critical
dimension of randomly pinned elastic manifolds, an expectation
which turns out to be correct both for the equilibrium properties
in the absence of a driving force and for the
fluctuationless ($T=0$) critical depinning transition at $f_c$ in the
presence of a driving force $f$. A renormalization group
$\epsilon$-expansion about $d_c=4$ then allows one to investigate
the various large scale properties of the pinned elastic manifold
\cite{Fisher}. An important feature of this renormalization group
analysis is the fact that even to lowest order one has to take
into account infinitely many variables, in practice by
renormalizing a function that is essentially the correlator
$\Delta [u({\bf z})-u({\bf z'})]$ of the random force (but see the
discussion in Ref.\ \onlinecite{NF-CDW} and in section
\ref{RG-formulation}).

The functional renormalization group (FRG) approach has been
successfully applied to both static equilibrium \cite{Fisher} and
fluctuationless ($T=0$) depinning problems
\cite{Narayan,Nattermann}. Recently, the FRG has also been applied
to the thermal creep problem \cite{Radzihovsky,Chauve,Mueller},
although its applicability is far more problematic. For forces
$f\ll f_c$, the motion proceeds either by activation or tunneling of
large segments between different metastable minima and thus the
dynamics is {\it intrinsically} non-perturbative: in particular, the
dynamic exponent $z$, which is shifted only slightly from its naive
value of two near the critical depinning transition, must be radically
modified in the creep regime. Nevertheless, it {\it appears}
(though, see section \ref{validity}) that for thermal depinning
one can handle this within the FRG and renormalize until a scale
is reached beyond which the disorder can be taken into account
perturbatively and the equation of motion can be solved. By
retracing back the FRG flow, one can then determine the average
velocity of the manifold. We will follow this approach here, but
return later in the paper to question its validity both for the
quantum and classical cases.

Formally, in both classical and quantum cases, the RG analysis of
creep involves the flow away from the static ($f = 0$) fixed-point
that controls the undriven pinned system under the action of a
small applied force $f$; under the RG transformation the
parameters describing the system flow in such a way that
eventually the effects of disorder can be neglected beyond a
certain length scale. But there is an important difference between
the classical and quantum cases: In the classical case
\cite{Chauve} one can restrict the analysis to the low-frequency
limit, i.e., including the friction term ($\eta\partial_t u$)
alone into the renormalization procedure is sufficient. This is a
consequence of the fact that, to exponential accuracy, classical
creep does not depend on the dynamics (e.g., inertial versus
dissipative) of the system. In the quantum case, it will turn out
that an analogous treatment leads to a spurious ``localization
transition'' where the average velocity of the manifold drops to
zero at a finite length scale $\sim L_c$, see
section~\ref{solution_of_rg_equations}. In order to carry out a
correct analysis one must include {\it all} the frequencies into
the renormalization group; specifically, we must make the
substitution
\begin{equation}
   \eta\partial_t u \rightarrow {\hat D} \otimes u
   \equiv \int_{-\infty}^t d\tau D(\tau)u(t-\tau)
   \equiv \eta\partial_t u
   +\sum\limits_{k=2}^{\infty}
   \eta^{(k)}\partial^{k}_t u.
   \label{dynamic_spectrum}
\end{equation}
It will turn out that, even if initially $\eta^{(k)}\equiv 0$ for
all $k\ge 2$, the dynamic parameters $\eta^{(k)}$ will grow
rapidly under the flow and become crucial; the renormalized
spectrum $D_l (\omega)$ is very different from the bare linear
spectrum $-i\omega \eta$.

It has been argued above that in order to investigate quantum
creep we can use the action (\ref{MSR_action}) with the correlator
given by (\ref{quantum_noise}). It is convenient to perform the
disorder average in (\ref{MSR_action}) and work in a frame moving
at the average velocity of the manifold, i.e., we substitute
$u({\bf z}, t)\rightarrow u ({\bf z}, t)+ vt$.  We also introduce
the above general dynamics ${\hat D}$; the resulting action,
averaged over both quantum fluctuations and random pinning, takes
the form
\begin{eqnarray}
   A &=& -\int\thinspace d^{d}z dt \thinspace
   iy \big({\hat D}\otimes u-c \partial^{2}_{\bf z} u\big)
   +\frac{1}{2}\int\thinspace d^{d}z dt dt^{\prime}\thinspace
   iy({\bf z}, t)\kappa (t-t^{\prime})iy({\bf z},t^{\prime})
   \label{complete_action}\\
   &+&\int\thinspace d^{d}z dt \thinspace
   iy\left (f-\eta v \right )+\frac{1}{2}\int\thinspace
   d^{d}zdtdt^{\prime}\thinspace
   iy({\bf z},t)\Delta\big[u({\bf z},t)-u({\bf z},t^{\prime})
   +v(t-t^{\prime})\big] iy({\bf z},t^{\prime}).
   \nonumber
\end{eqnarray}
The correlation functions corresponding to the quadratic action in
the absence of pinning are given by
\begin{eqnarray}
   \overline{u({\bf k},\omega) u({\bf k}^{\prime},\omega^{\prime})}
   &=&\frac{\kappa(\omega)}{{\left|c{\bf k}^{2}+D(\omega)\right|}^{2}}
   \delta(\omega+\omega^{\prime})
   \delta({\bf k}+{\bf k}^{\prime})
   \equiv C ({\bf k},\omega) \delta(\omega+\omega^{\prime})
   \delta({\bf k}+{\bf k}^{\prime}),
   \label{definition_c}\\
   \overline{u({\bf k},\omega) iy ({\bf k}^{\prime},\omega^{\prime})}
   &=& \frac{1}{c{\bf k}^{2}+D(\omega )} \delta(\omega+\omega^{\prime})
   \delta({\bf k}+{\bf k}^{\prime})
   \equiv R({\bf k}, \omega) \delta (\omega+\omega^{\prime})
   \delta({\bf k}+{\bf k}^{\prime}),
   \label{definition_r}
\end{eqnarray}
with the usual definition of the Fourier transform
\begin{equation}
   f({\bf k}, \omega) \equiv \int\thinspace d^{d}z dt
   \thinspace f({\bf z},t) \exp\left(-i{\bf k}{\bf z}
   + i\omega t\right)
   \label{FT}
\end{equation}
and $D(\omega) \approx \eta (-i\omega)$ for small $\omega$.
Because of the fluctuation dissipation theorem, we have
\begin{equation}
   \kappa(\omega)
   = -\hbar\, {\rm Im}D(\omega)\>\coth\frac{\hbar\omega}{2T}.
   \label{kappa_omega}
\end{equation}

In the regime of interest for quantum creep, the motion of the
elastic manifold can be represented as jumps between different
metastable states and the lifetime of each metastable state is
large. It is thus possible to extract some information about the
dynamic properties of the system, in particular, the average
velocity $v$, from {\it static} equations if we cutoff the
renormalization group flow at the relevant length scale, here,
$L_{f}\sim L_c {\left (f_{c}/f\right )}^{1/{(2-\zeta)}}$, see
section~\ref{scaling_arguments} (a similar situation arises in the
conventional theory of the decay of metastable states: although
the original problem is formally non-equilibrium, the system is
trapped for a long time in the local equilibrium state
corresponding to a local minimum of the free energy; one can
approximate the partition function of the system by that
calculated assuming quasi-equilibrium and then find the flow of
the probability out of the local potential well). This is why we
could use the Euclidean action instead of the full quantum
mechanical action in section~\ref{scaling_arguments}. For values
of the RG variable $l$ smaller than $\ln(L_f/L_c)$ we then can use
the appropriate static versions of RG equations.

The RG flow equations are obtained by integrating over fast modes
in the effective action (\ref{complete_action}) with the last two
terms in the action treated as perturbations about the quadratic
action. We write all the equations, except those for the
correlator $\Delta_l(u)$, the temperature $T_l$, and the Planck
constant $\hbar_l$, at finite velocity. We will use the subscript
$l$ in order to distinguish renormalized quantities at length
scale $L=\Lambda^{-1} e^l$ from bare quantities without
subscripts, e.g., $\eta=\eta_0$, $T = T_0$, $\hbar = \hbar_0$,
$\Delta (u) = \Delta_0 (u)$, etc. Introducing the large momentum cutoff
$\Lambda$ we obtain to leading non-trivial order in $\Delta_l$,
\begin{eqnarray}
   \partial_l\Delta_{l} (u) &=&
   \left(4-d-2\zeta\right )\Delta_{l}(u)
   +\zeta u\Delta_{l}^{\prime}(u)+C_{l}^{>}(\Lambda ,t=0)
   \Delta^{\prime\prime}_{l}(u)\nonumber\\
   &+&I\Delta_{l}^{\prime\prime}(u)
   \big[\Delta_{l}(0)-\Delta_{l}(u)\big]
   -I{\Delta_{l}^{\prime}}^{2} (u), \label{correlator_rg}\\
   \partial_l{\tilde f}_{l} &=&
   (2-2\zeta){\tilde f}_{l}
   +\int\thinspace dt R^{>}_{l}(\Lambda ,t)
   \Delta_{l}^{\prime}(v_l t),
   \label{force_rg}\\
   \partial_l v_{l}&=&(z-\zeta)\>v_{l},
   \label{velocity_rg}\\
   \partial_l \kappa_{l}(t) &=&
   (4-d-2\zeta)\kappa_{l}(t)
   +z t \partial_t \kappa_{l}(t)
   -C_{l}^{>}( \Lambda,t)\Delta^{\prime\prime}_{l}(v_l t),
   \label{noise_rg}\\
   \partial_l D_{l}(\omega) &=&
   2 D_{l}(\omega)-z\omega \partial_\omega
   D_{l}(\omega)-\int \thinspace\frac{d\omega^{\prime}}{2\pi}
   \frac{1}{v_l}{\hat \Delta}_{l}^{\prime\prime}
   \left(\omega^{\prime}/v_l\right)
   \left[ R^{>}_{l}(\Lambda ,\omega^{\prime})
   -R^{>}_{l}(\Lambda,\omega+\omega^{\prime})\right],
   \label{dynamic_rg}\\
   \partial_l T_l &=& (2 - d - 2\zeta)\>T_l,
   \label{temperature_renormalization}\\
   \partial_l \hbar_l &=&
   (2-d-2\zeta-z)\>\hbar_l,
   \label{hbar_renormalization}
\end{eqnarray}
with $I \equiv A_{d}\Lambda^d/c^{2}\Lambda^4$ defined in terms of
the surface area $A_{d}$ of a $d$-dimensional unit sphere. Both
the dynamic exponent $z=z(l)$ and the roughness exponent
$\zeta(l)$ are at our disposal to adjust for convenience; it will
generally be most useful to choose $\zeta$ to be the
$l$-independent value giving rise to a well behaved fixed point
function $\Delta^\ast(u)$ in the absence of fluctuations or drive.
How best to adjust $z(l)$ we reserve for later; conventionally it
would be chosen to fix the coefficient in the low-frequency part
of $D_l(\omega)$, in the dissipative situation studied here, of
$\eta_l$. The
shell-restricted correlation functions $C^{>}(\Lambda,t)$ and
$R^{>}(\Lambda,t)$ are given by
\begin{equation}
   C^{>}_{l}(\Lambda ,t)
   = \frac{A_{d}\Lambda^{d}}{{\left (2\pi \right)}^{d}}
   \int \thinspace \frac{d\omega}{2\pi} e^{-i\omega t}
   C(\Lambda ,\omega)
   \label{equation_c}
\end{equation}
and
\begin{equation}
   R^{>}_{l}(\Lambda ,t)
   = \frac{A_{d}\Lambda^{d}}{{\left(2\pi \right)}^{d}}
   \int\thinspace \frac{d\omega}{2\pi}
   e^{-i\omega t} R(\Lambda,\omega);
   \label{definition_r_>}
\end{equation}
here, $C(\Lambda,\omega)$ and $R(\Lambda ,\omega)$ are defined via
(\ref{definition_c}) and (\ref{definition_r}) with $D(\omega)$
substituted by $D_{l}(\omega)$. Also, we define the coefficient
\begin{equation}
   \Gamma_l\equiv C_{l}^{>}(\Lambda ,t=0),
   \label{Gamma}
\end{equation}
which will appear frequently in the following. We chose an initial
dynamic spectrum of the form
\begin{equation}
   D_{0}(\omega) = -i\eta \omega + \rho\omega^2.
   \label{initial_condition_1}
\end{equation}
The excess of the applied force over that needed to sustain the
motion of an unpinned system with dissipative coefficient $\eta_l$
is
\begin{equation}
   {\tilde f}_{l} = f_{l} - \eta_l v_l.
   \label{tilde_f}
\end{equation}
Above, we have defined the Fourier transform of the second
derivative $\Delta''(u) \equiv \partial_u^2 \Delta$ by ${\hat
\Delta}^{\prime\prime}(p) \equiv \int du\, e^{ipu}\Delta''(u)$, so
that
\begin{equation}
   \hat{{\Delta}}_{l}^{\prime\prime}(\omega /v)
   \equiv v \int_{-\infty}^{+\infty} \thinspace dt
   e^{i\omega t} \Delta^{\prime\prime}_{l}(vt);
   \label{Dpp}
\end{equation}
this quantity plays a crucial role in the dynamic renormalization.
Eqs.~(\ref{correlator_rg}), (\ref{force_rg}), (\ref{noise_rg}),
(\ref{velocity_rg}), and (\ref{temperature_renormalization}) have
been obtained before in the discussion of classical creep
\cite{Chauve}; they are the same in the quantum case, while
Eq.~(\ref{dynamic_rg}) is different, however. We now show how it
can be obtained and in what respect it differs from its classical
analog.

After averaging the disorder term over the fast modes one obtains
the following feedback $\delta A$ to the term $-\int\thinspace
d^{d}zdt\, iy({\bf z},t){\hat D}_{l}\otimes u$ in the action,
\begin{eqnarray}
   \delta A =
   -\int d^{D}zdt dt^{\prime}\thinspace iy({\bf z},t)
   R^{>}_{l}\left(\Lambda , t-t^{\prime}\right)
   \left [u({\bf z},t) - u({\bf z},t^{\prime})\right]
   \Delta^{\prime\prime}_{l}\left[v(t-t')\right]dl.
   \label{correction_to_dynamics}
\end{eqnarray}
Writing the functions $\Delta_{l}(v \tau)$ and $R^{>}_{l}(\Lambda
,\tau)$ as Fourier integrals we obtain Eq.~(\ref{dynamic_rg}),
which explicitly includes the full dependence of the displacement
field $u({\bf z},t)$ on time, i.e., it is nonperturbative in
frequency. In the classical analysis of Ref.~\onlinecite{Chauve}
it was assumed that the low-frequency limit of the RG equations is
valid for all frequencies, i.e., for $\omega \rightarrow 0$ the
only important term is that containing $\eta_{l} \partial_t u$
(corresponding to $-\eta_{l} i\omega u(\omega )$ in Fourier
space). The equation analogous to (\ref{dynamic_rg}) then is given
by its low-frequency limit, i.e., instead of the full dependence
$u({\bf z}, t) - u({\bf z}, t^{\prime})$ one considers only the
first term of its Taylor expansion in $t-t^{\prime}$. It was also
assumed in Ref.~\onlinecite{Chauve} that the response function
$R^{>}_{l}(\Lambda, t)$ and the correlation function
$C_{l}^{>}(\Lambda ,t)$ depend only on the frictional coefficient
$\eta_{l}$. In the classical limit this approach leads to
reasonable results. By contrast, ignoring the frequency dependence
of the dynamics in the quantum case leads to a spurious
``localization transition'', implying a zero average velocity of
the manifold below a small but non-zero driving force $f$ and in
the presence of quantum fluctuations $\hbar \ne 0$. We will
discuss this issue in section~\ref{solution_of_rg_equations}.

In the case of a purely dissipative dynamics there is another
problem that occurs even at the initial stage of the
renormalization: the integral in the expression for
$C^{>}_{l}(\Lambda ,t = 0)$ diverges at large frequencies, see
Eqs.~(\ref{equation_c}) and (\ref{definition_c}). For large
$\omega$ the noise $\kappa_l(\omega)$ (see (\ref{quantum_noise}))
and the dynamic spectrum $D_l(\omega)$ are both proportional to
$\omega$ for a dissipative dynamics and, consequently,
$C_{l}^{>}(\Lambda,t = 0)$ diverges logarithmically as $\int
\thinspace d\omega /\omega$ at large frequencies, see
(\ref{definition_c}). This is not unexpected: it is just such a
logarithmic frequency dependence that can cause localization in
models with a single degree of freedom coupled to a bath that
provides a linear friction \cite{Leggettetal}; thus integrating
out all the frequencies at once is problematic even for short
wavelength deformations of the manifold. In reality, we expect
that an inertial term $\rho\partial^2_t u$ or some other frequency
dependence describing the small scale dynamics will provide a
cutoff at high frequencies (at $\omega_0 \sim \eta/\rho $ for the
inertial case); alternatively one could introduce a sharp cutoff
by hand. We will consider both possibilities later, noting now
that how this is done will affect the results far more than one
might expect.

\subsection{Structure of the Quantum RG Flow}
\label{solution_of_rg_equations}

The main goal of the remainder of this section is to analyze the
system of RG equations derived in section
\ref{derivation_rg_equations} in the limit of small driving
forces. We first discuss the important features and then provide a
more detailed analysis in the following section. First, let us
analyze the renormalization (\ref{correlator_rg}) of the
force-force correlator. In the absence of quantum and thermal
fluctuations (i.e., $\kappa (\omega) = 0$ for all $\omega$), the
correlation function $C_{l}^{>}(\Lambda,t=0)$ is zero. In this
case, the nonlinearities in the flow equation of the function
$\Delta_l(u)$ cause it to become nonanalytic on a finite length
scale, even if the bare correlator $\Delta_0 (u)$ is analytic, see
Ref.\ \onlinecite{Fisher}. This is easily seen by differentiating
Eq.~(\ref{correlator_rg}) twice with respect to $u$ and
substituting $u=0$, resulting in an equation for the evolution of
the quantity $\Delta_l^{\prime\prime} (0)$ alone. The simple
autonomous flow equation for $\Delta_{l}^{\prime\prime}(0)$ leads
to a divergence at a finite scale $l_c \sim \epsilon^{-1}
\ln[c^2/\Lambda^{d-4}\Delta_0^{\prime\prime}(0)]$, producing the
Larkin- or pinning length $L_c \simeq [c^2/\Delta^{\prime\prime}
(0)]^{1/(4-d)}\sim [c^2\xi^2/\Delta(0)]^{1/(4-d)}$ where
collective pinning goes over into strong pinning. Beyond $L_c$,
the perturbative description breaks down as multiple competing
minima appear in the pinning energy landscape. The infinite second
derivative suggests that the function $\Delta_l(u)$ will have a
discontinuous first derivative at $u=0$, i.e., $\Delta^{\prime}_l
(+0) = - \Delta^{\prime}_l (-0) \ne 0$. On length scales shorter
that $L_c$, the smoothness of $\Delta_l(u)$ reflects the smooth
reversible evolution of a segment as it is pulled by its
neighboring regions. But on larger length scales the internal
deformations of the segment will cause it to jump discontinuously
and irreversibly from one metastable configuration to another ---
this is reflected in the discontinuity in $\Delta^\prime_l$. At
the scale $L_c$ the force correlator essentially has reached its
fixed point shape with a height $\Delta^\ast(0)\sim
(c\Lambda^2)^2\xi^2\Lambda^{-d} e^{-2\zeta l_c}$ and a width
$\xi^\ast \sim \xi e^{-\zeta l_c}$, resulting in a cusp
$|{\Delta^\ast}^\prime(0+)| \sim (c\Lambda^2)^2 \xi \Lambda^{-d}
e^{-\zeta l_c}$, see appendix \ref{appendix1}.

In the presence of small quantum or thermal fluctuations the
coefficient $\Gamma_l \equiv C_{l}^{>}(\Lambda,t=0)$ becomes
nonzero and this leads to a {\it smearing} of the cusp in
$\Delta_l(u)$ \cite{Radzihovsky,Chauve,Balents}. The derivative
$\Delta^{\prime}_l(0)$ of the correlator at the origin is zero for
$C^{>}(\Lambda,t=0) \ne 0$ but changes rapidly in a boundary layer
around the origin. The cusp that was present in the absence of
fluctuations is smeared over a region $u_{\rm smear}$ which can be
estimated by comparing \cite{Chauve} the terms $\Gamma_l \left
|\Delta_{l}^{\prime\prime}(0)\right| \sim \Gamma_l \left
|\Delta_l^{\prime} (u\sim u_{\rm smear}) \right |/{u_{\rm smear}}$
and $[\Lambda^d/(c\Lambda^2)^2]{\Delta_l^{\prime}}^{2} (u)$ in
(\ref{correlator_rg}). The derivative $\Delta_l^{\prime} (u >
u_{\rm smear})$ approaches the fixed point value
${\Delta^{*}}^{\prime}(+0)$ found in the absence of fluctuations
and we obtain the boundary width $u_{\rm smear} \sim \Gamma_l
(c\Lambda^2)^2/\Lambda^d |{\Delta^\ast}^\prime(0+)|$. The
curvature $\Delta_l^{\prime\prime}(0) \sim -[\Lambda^d/(c
\Lambda^2)^2]|{\Delta^\ast}^\prime(0+)|^2/\Gamma_l$ then diverges
as the fluctuations renormalize to zero on large scales.

Physically, $u_{\rm smear}(l)$ can be understood in terms of the
{\it equilibrium} response of a segment of size $L=\Lambda^{-1}
e^l$ to the motion of its neighboring regions. Usually, a small
displacement of neighboring regions will cause only a small
readjustment of the segment of interest within its local energy
minimum. The exceptions to this occur when the minimum energy
configuration of the segment jumps from one configuration to
another as its neighboring regions are slightly displaced: it is
these jumps that give rise to the cusp in $\Delta(u)$ in the
absence of fluctuations. At non-zero temperature or in the
presence of quantum fluctuations, the behavior will not change
much except near these jumps, where there will be a range of
positions of the neighboring regions over which the segment of
concern will have a non-negligible probability to be in {\it
either} of two distinct configurations. This will result in a
smearing of the cusp in $\Delta_l (u)$ over the scale $u_{\rm
smear}(l)$ of neighboring region displacements on which this split
probability typically occurs. On large scales, the fact that the
energy scale grows as $L^\theta$ means that it is much less likely
that the position of a segment will fluctuate between two energy
minima. This is reflected in the renormalization towards zero of
$\Gamma_l$ and the concomitant flow towards zero of the smearing
scale $u_{\rm smear}\propto\Gamma_l$. Summarizing, the function
$\Delta_l (u)$, whose renormalization is given through equation
(\ref{correlator_rg}), develops under the RG in the following way:
At the scale $L_c$, $\Delta_l(u)$ is close to its fluctuationless
fixed-point function $\Delta^\ast(u)$, with fluctuations affecting its
behavior only in the vicinity of the point $u=0$ via a smearing of
the discontinuity in $\Delta'_l$ in a boundary layer whose size is
controlled by $\Gamma_l$. Once we know how the function $\Delta_l
(u)$ evolves, we can substitute it into the other RG equations and
see how other quantities renormalize under the RG flow.

In addition to $u_{\rm smear}$, there are two other important
displacement scales: the characteristic scale $\xi^{*}$ of the
fixed-point correlator $\Delta^\ast(u)$ and the scale $u_{\rm
vel}$ associated with the velocity $u_{\rm vel}\sim \eta_l
v_l/c\Lambda^2$. The latter is the product of the velocity $v_l$
and the characteristic timescale $\eta_l/c\Lambda^2$ of the
low-frequency part of the response function $R^{>}_l(\Lambda,t)$ at
wavelengths of the order of the cutoff $\Lambda^{-1}$. At scales
somewhat larger than the Larkin length $L_c \propto \Lambda^{-1}
e^{l_c}$,
\begin{equation}
   u_{\rm vel}\ll u_{\rm smear}\ll \xi^{*}.
   \label{u_vel}
\end{equation}
This is a consequence of the fact that $v_l$ is {\it
exponentially} small ($\propto \exp\left\{-\left[S(f)/\hbar
\right]^{\alpha} \right \}$) in $\hbar$, while $u_{\rm smear}$
proportional to a {\it power} of $\hbar_l$ and $\xi^{*}$ is a
static quantity that does not depend on $\hbar_l$ (in the
classical case the relevant displacement scales obey the same
relation with the role of $\hbar$ played by the temperature $T$).
During the RG flow, $u_{\rm smear}$ decreases gradually with
decreasing $\hbar_l$ whereas $u_{\rm vel}$ increases rapidly due
to the sharp increase in the viscosity $\eta_l$.

Eventually at some scale both $u_{\rm smear}$ and $u_{\rm vel}$
are of the same order.  At this scale, which is the crossover
scale $L_f$ appearing in the scaling arguments for creep, the
functions $\Delta^{\prime}_l (vt)$ and $\Delta^{\prime\prime}_l
(vt)$ are roughly given by ${\Delta^{*}}^{\prime} (+0)$ and
${\Delta^{*}}^{\prime\prime}(+0)$ on the time scales that dominate
in Eqs.~(\ref{force_rg}), (\ref{noise_rg}), and
(\ref{dynamic_rg}). At even longer scales, $u_{\rm vel}$
eventually becomes comparable to $\xi^{*}$, see Ref.\
\onlinecite{Chauve}. Beyond this scale the randomness is
effectively smoothed by the motion of the manifold and can be
neglected or treated perturbatively.

In analyzing the renormalization of other quantities by the random
pinning forces, we thus see that there are two important regimes:
\begin{equation}
   {\it i})\quad {u_{\rm vel}} < {u_{\rm smear}},
   \quad {\rm the\  nucleation\ regime},
   \label{nuc_regime}
\end{equation}
where we can approximate $\Delta_l^{\prime} (vt)$ by
$\Delta_l^{\prime} (0) = 0$ and $\Delta_l^{\prime\prime} (vt)$ by
$\Delta_l^{\prime\prime} (0)$. And
\begin{equation}
   {\it ii}) \quad {u_{\rm smear} < u_{\rm vel}}< \xi^{*},
   \quad {\rm the\ depinning\ regime};
   \label{dep_regime}
\end{equation}
it was argued \cite{Chauve} that this latter regime resembles that
of the critical depinning transition \cite{Narayan} in the absence
of fluctuations; we can then approximate $\Delta_l^{\prime} (vt)$
by ${\Delta_l^{*}}^{\prime} (+0)$ and $\Delta_l^{\prime\prime}
(vt)$ by ${\Delta_l^{*}}^{\prime\prime} (+0)$. These two regimes
are separated by the scale $L_f$ discussed in
section~\ref{scaling_arguments}. We will now show how the scale
$L_{f}\sim L_{c} {\left (f_{c}/f\right )}^{{1}/{(2-\zeta)}}$
naturally appears in the solution of the RG equations and
separates the two regimes.

In order to do so we concentrate on the force equation
(\ref{force_rg}) which involves the slope $\Delta_l^\prime(v_l t)$
of the disorder correlator near the origin. The latter is small
and vanishes at $u=0$ for $l<l_c$, hence ${\tilde f}_{l} =
e^{(2-\zeta)l} {\tilde f}$ grows exponentially (since the velocity
$v_l$ of the manifold is exponentially small we can neglect the
term $\eta_l v_{l}$ in the equation for ${\tilde f}_{l}$). In the
absence of fluctuations the slope $\Delta_l^\prime(v_l t) \sim
\Delta^{\ast\prime}(0+)$ rapidly turns on as the correlator forms
a cusp at $l_c$; if the disorder term $(\Lambda^d/c\Lambda^2)|
\Delta^{\ast\prime}(0+)|$ overcompensates the scaling term
$(2-\zeta )e^{(2-\zeta )l_{c}}{\tilde f}$, the force will start
renormalizing to zero while in the opposite case it will continue
to increase. Balancing the two terms we find the critical force
density $f_c \sim c \xi / L_c^2$.

Fluctuations smearing the correlator on the scale $u_{\rm smear}$
soften and delay the depinning transition until the growing width
$u_{\rm vel}$ of the response function encloses the emerging cusp
in $\Delta_l(u)$ at $l_f$, $u_{\rm vel}(l_f) \sim u_{\rm
smear}(l_f)$. Below $l_f$ we have $\Delta_l^\prime(v_l t) \approx
\Delta_l^\prime(0) = 0$ and the force increases exponentially
${\tilde f}_{l} = e^{(2-\zeta)l} {\tilde f}$; starting with a
small force $f \ll f_c$, depinning occurs as the cusp emerges at
$u_{\rm smear}(l_f) \sim u_{\rm vel}(l_f)$. Replacing again
$\Delta_l^\prime(v_l t) \approx \Delta^{\ast\prime}(0+)$ in
(\ref{force_rg}) we obtain the depinning condition ${\tilde
f}_{l_f} \sim (\Lambda^d/ c\Lambda^2)| \Delta^{\ast\prime}(0+)|
\sim f_c e^{(2-\zeta)l_c}$; the depinning scale $l_f$ then relates
to the force $f$ via $(2-\zeta)(l_f-l_c) \sim \ln(f_c/f)$ or
\begin{equation}
   L_f \sim L_c {\left(f_c/f\right)}^{{1}/{( 2-\zeta )}}.
   \label{L_f_rg}
\end{equation}
We thus see that the characteristic length scale for creep --- the
size $L_f$ of the segments that can jump to lower the energy ---
appears naturally within the RG framework. We emphasize, however,
that $L_f$ is entirely determined by {\it static} properties; so
far we have not had to address the crucial issue of dynamic
renormalizations --- to these we now turn.

The crucial quantity needed for the dynamic renormalization is
$\Delta_l''(0)$, which can be obtained from
Eq.~(\ref{correlator_rg}) by comparing ${\Gamma_l}\Delta^{\prime
\prime}_{l}(u)$ with the nonlinear terms evaluated at the origin
$u = 0$. This yields \cite{Chauve}
\begin{equation}
   -{\Delta_{l}}^{\prime\prime}(0) \approx
   \frac{1}{\Gamma_l} \frac{\Lambda^{d}}
   {{\left({c\Lambda}^{2}\right)}^{2}}
   {\Delta_{l}^{\prime}(+0)}^{2}
   \label{smearing}
\end{equation}
(see appendix \ref{appendix1}) and enables us to find the
renormalization of the viscosity $\eta_l$. Since we are primarily
interested in calculating the velocity to exponential accuracy,
the calculations simplify significantly. If we do {\it not}
renormalize time scales explicitly, i.e., choosing $z(l)=0$, the
velocity $v$ is given approximately by the viscosity at the scale
$L_f$,
\begin{equation}
   \ln v\approx \ln\frac{1}{\eta_{l_{f}}}.
   \label{relation}
\end{equation}
The validity of this condition to exponential accuracy is due to
the following argument: At the scale $L_{\rm dyn}$ where $u_{\rm
vel}\sim\eta_l v_l/c\Lambda^2 \sim\xi^{*}$ we have a velocity $v_l
\propto 1/\eta_l$ as the effects of pinning are negligible beyond
this scale. On intermediate scales, i.e., between $L_f$ and
$L_{\rm dyn}$ both $\eta_l$ and $v_l$ renormalize but {\it not}
exponentially; the only exponentially large renormalization
originates from scales between $L_c$ and $L_f$. Thus, to
exponential accuracy, we can ignore the renormalization on scales
larger than $L_f$ (and smaller than $L_c$) and justify
(\ref{relation}) (note that, in general, only such dynamic
quantities as $\eta_l$ and $v_l$ are exponentially renormalized,
i.e., proportional to $\exp[\pm {\rm const.} L_f^{\tilde
\alpha}]$, with ${\tilde\alpha}$ a positive constant).
\cite{v-scales}

The crucial equations for calculating the renormalized velocity
are Eqs.~(\ref{noise_rg}) and (\ref{dynamic_rg}). Substituting
$v_l$ by zero in these equations and performing the Fourier
transformation in Eq.~(\ref{noise_rg}) we obtain the two equations
\begin{equation}
   \partial_l\kappa_{l}(\omega) =
   \left (4-d-2\zeta-z\right )\kappa_{l}(\omega)
   -z\omega \partial_\omega\kappa_{l}(\omega)
   -\frac{A_{d}\Lambda^{d}}{{\left(2\pi\right)}^{d}}
   \Delta_{l}^{\prime\prime}(0)\frac{\kappa (\omega )}
   {{\left|c\Lambda^{2}+D_{l}(\omega )\right|}^{2}},
   \label{new_noise_rg}
\end{equation}
\begin{equation}
   \partial_l D_{l}(\omega ) =
   2 D_{l}(\omega )-z\omega \partial_\omega
   D_{l}(\omega)-\frac{A_{d}\Lambda^{d}}{{\left(2\pi\right)}^{d}}
   \Delta_{l}^{\prime\prime}(0)\frac{D_{l}(\omega)}
   {c\Lambda^2[c\Lambda^{2}+D_{l}(\omega)]}.
   \label{new_dynamic_rg}
\end{equation}

Note that Eqs.~(\ref{new_noise_rg}) and (\ref{new_dynamic_rg}) are
very similar except for the trivial scaling parts. This is the
consequence of the quantum fluctuation--dissipation theorem
\cite{Landau} (FDT),
\begin{equation}
   \kappa_l (\omega)
   = \frac{\hbar_l {\rm Im}\left[R^{>}_l(\Lambda,\omega)\right]}
   {{\left|R^{>}_l(\Lambda,\omega )\right|}^2}
   \coth\frac{\hbar_l\omega}{2T_l},
   \label{fdt}
\end{equation}
which is valid in the quasi-equilibrium situation produced by the
long time scales associated with the creep motion. If the relation
(\ref{fdt}) is satisfied in the bare system ($l=0$), then
Eqs.~(\ref{new_noise_rg}) and (\ref{new_dynamic_rg}) guarantee its
validity for any $l$. Strictly speaking the FDT is not applicable
for $v \ne 0$, but its appearance here is understood to be
consistent with our assumption of being close to local equilibrium
and our use of the approximation $v = 0$ in the RG equations for
$L<L_f$. Using Eqs.~(\ref{definition_r}), (\ref{definition_r_>}),
and (\ref{fdt}) one can see that the expressions
(\ref{new_noise_rg}) and (\ref{new_dynamic_rg}) are identical up
to scaling terms.

Substituting Eq.~(\ref{smearing}) into (\ref{new_dynamic_rg}) we
obtain
\begin{equation}
   \partial_l D_{l}(\omega ) =
   2 D_{l}(\omega)-z\omega \partial_\omega
   D_{l}(\omega )+\frac{A_{d}\Lambda^{d}}{{\left(2\pi\right)}^{d}}
   \frac{\Lambda^{d}}{{\left (c\Lambda^{2}\right )}^{3}}
   \frac{{{\Delta_{l}}^{\prime}(+ 0)}^{2}}{\Gamma_l}
   \frac{D_{l}(\omega )}{
   \left[c\Lambda^{2}+D_{l}(\omega )\right]}.
   \label{single_rg_equation}
\end{equation}
Note that $\Gamma_l$ and $R_{l}^{>}(\Lambda,t = 0)$ are functions
of $l$ via their dependence on $D_l(\omega)$. Using
Eqs.~(\ref{definition_c}), (\ref{definition_r}), and (\ref{fdt})
we can write $\Gamma_l$ in the form
\begin{equation}
   \Gamma_l=-\frac{A_d\Lambda^d}{{\left(2\pi\right)}^d}
   {\int\limits_{-\infty}^{+\infty}\thinspace
   \frac{d\omega}{2\pi}\frac{\hbar_{l}{\rm Im}
   \left[D_{l}(\omega)\right]
   \coth(\hbar_l\omega/2T_l)}
   {{\left|c\Lambda^{2}+D_{l}(\omega)\right |}^{2}}}.
   \label{expression_for_c}
\end{equation}
In a nearly static system $\hbar_{l}$ and temperature $T_{l}$
renormalize to zero in a trivial way, see Eqs.\
(\ref{temperature_renormalization}) and
(\ref{hbar_renormalization}), and the elasticity $c$ is not
renormalized at all, $\partial_l c=0$. This is a consequence of
the statistical tilt symmetry ($u\rightarrow u + {\bf b}\cdot{\bf
z}$, with ${\bf b}$ an arbitrary constant vector) of the action
(\ref{MSR_action}).

We now concentrate on the zero temperature limit where the
correlation function $\Gamma_l=C_{l}^{>}(\Lambda ,t=0)$ can be
written in the form (cf., (\ref{definition_r}) and (\ref{fdt}))
\begin{equation}
   \Gamma_l=\frac{A_d\Lambda^d}{{\left (2\pi\right )}^d}
   \int\limits_{0}^{\infty}\thinspace \frac{d\omega}{2\pi}
   \frac{2\hbar_l}{c\Lambda^{2} + D_{l}(i\omega )}.
   \label{new_expression_for_c}
\end{equation}
This is a consequence of the identity \cite{Landau}
\begin{equation}
   \int\limits_{0}^{\infty}\thinspace{d\omega}\thinspace
   R^{>}_{l}(\Lambda,i\omega )
   =\int\limits_{0}^{\infty}\thinspace{d\omega}\thinspace
   {\rm Im} [R^{>}_{l}(\Lambda,\omega)]
   \label{identity}
\end{equation}
following from the analytic structure of the correlator $R_l$ as
implied by causality. It is convenient to work within the
imaginary time formalism and substitute $\omega \rightarrow
i\omega$. This substitution transforms the dynamic spectrum into
$D_l( i\omega) = \eta\omega + \sum_{k\ge 2}\eta^{(k)}\omega^{k}$
with the advantage that $D_l(i\omega)$ is real and nonnegative for
positive $\omega$. The calculation of the velocity $v$ of the
driven elastic manifold reduces to the problem of solving the
equation
\begin{eqnarray}
   \partial_l D_{l}(i\omega ) &=&
   2 D_{l}(i\omega)-z\omega\partial_\omega D_{l}(i\omega )
   \label{imaginary_rg_equation}\\
   &+&\frac{A_{d}\Lambda^{d}}{{\left(2\pi\right)}^{d}}
   \frac{\Lambda^{d}}{{\left (c\Lambda^{2}\right)}^{3}}
   \frac{{{\Delta_{l}}^{\prime}(+ 0)}^{2}}{\Gamma_l}
   \frac{D_{l}(i\omega )}
   {\left[c\Lambda^{2}+D_{l}(i\omega)\right]},
   \nonumber
\end{eqnarray}
with the initial condition $D_{0}(i\omega )=\eta|\omega|+\rho
\omega^2$ (valid for any $\omega$) and calculating the
renormalized low-frequency viscosity $\eta_{l}=
\partial_{\omega}D_{l}(i\omega )$, $\omega\rightarrow 0$; having
found $\eta_{l}$ we can determine the velocity $v$ using the
relation (\ref{relation}). Note that if $D_0 (i\omega )=D_0
(-i\omega )$ then this property is preserved under the RG
transformation.

On very short scales $l<l_c$ the disorder-dependent term on the
right-hand side of (\ref{imaginary_rg_equation}) can be neglected.
On intermediate scales $l_c<l<l_f$ we can substitute
${\Delta_l}^{\prime}(+0)$ by its fixed-point value which is
related to the bare potential through (see appendix
\ref{appendix1})
\begin{equation}
   {{\Delta^{*}}^{\prime} (+ 0)}^{2}\simeq
   \epsilon e^{(\epsilon - 2\zeta)l_c}
   \frac{(c\Lambda^2)^2}{\Lambda^d} \Delta (0)
   \sim \frac{(c\Lambda^2)^4}{\Lambda^{2d}} \xi^2  e^{-2\zeta l_c},
   \label{useful_relation}
\end{equation}
where we have used the relation $\Delta(0) \sim c^2 \xi^2
(\Lambda/e^{l_c})^\epsilon $ in the last equation.

\subsection{Naive RG and ``Localization'' Transition}
\label{loc_tr}

Before embarking on the complete analysis it is instructive to see
what happens if we simply keep the leading low-frequency form of
$D(i\omega)$ as is conventionally done in dynamic renormalization
group calculations. To do this we substitute the Ansatz
$D_{l}(i\omega) = \eta_{l}|\omega|$ for {\it all} frequencies into
(\ref{new_expression_for_c}). Calculating the integral we easily
obtain $\Gamma_l \sim\left ({\hbar_{l}{\Lambda^d}}/{\eta_l}
\right)\ln \left (\eta_l{\bar\omega}/c\Lambda^2 \right)$, where we
have introduced a high-frequency cutoff ${\bar\omega}$.
Substituting this expression into (\ref{imaginary_rg_equation})
and approximating
\begin{equation}
   D_l(i\omega)/{\left(c\Lambda^{2}+D_l(i\omega )\right)}
   \rightarrow D_l( i\omega)/c\Lambda ^2
   =\eta_l\omega/c\Lambda^2,
\end{equation}
i.e., assuming that the low-frequency asymptotics $D_{l}(i\omega)
= \eta_{l}\omega$ is valid for any $\omega$ and $l$, we obtain the
equation for the renormalized friction coefficient
\begin{equation}
   \partial_l\eta_l
   \sim \frac{S_{\eta}}{\hbar\eta}e^{-(d+2\zeta)l_c}\, \eta_l^2,
   \label{eta_l}
\end{equation}
with $S_\eta \sim \eta \xi^2 L_c^d$.  Unfortunately, the behavior
of this equation is pathological: one can see that it would imply
a divergence of $\eta_{l}$ at a {\it finite} length scale. This
would presumably mean that the velocity goes to zero in the presence of a
nonzero force and quantum fluctuations, a
result that appears to be implausible. Of course, what one must
check in any situation where some parameter in an effective action
diverges under the RG flow is, whether this is due to an
unphysical restriction of the space of relevant parameters, a
breakdown in whatever approximations that have been made in
deriving the RG flow, or some other effect. In our case, it will
turn out that the renormalization of the whole frequency spectrum
is very important. This should be contrasted with the classical
case for which one can obtain the result $v\propto\exp[-U(f)/T]$
by considering the low-frequency limit only \cite{Chauve}.

\subsection{RG Analysis of Dynamic Response}
\label{full_analysis}

\begin{figure}
\centerline{\epsfxsize = 6.0cm \epsfbox{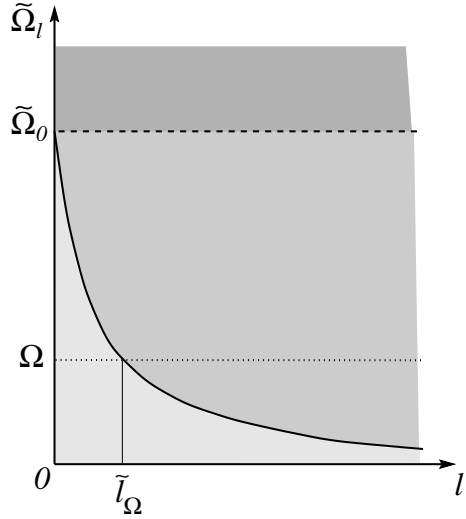}}
\vspace{0.3truecm} \caption{Different regimes relevant in the
integration of the dynamical equation (\ref{equation_for_e}): For
$\Omega < \tilde{\Omega}_0$ the integration is split into two
regimes $0<l<\tilde{l}_\Omega$ and $\tilde{l}_\Omega < l$, while
for $\Omega > \tilde{\Omega}_0$ the integration involves only one
regime.} \label{fig:Omega_l}
\end{figure}

We now turn to the analysis of the RG flow equations
(\ref{imaginary_rg_equation}) and (\ref{useful_relation}). In
order to find the flow of the function $D_l(\omega)$ it is
convenient to make the unconventional choice $z(l)=0$ and allow
$\eta_l$ to change arbitrarily. If one is more comfortable with a
flowing dynamical exponent $z(l)$, one can work more generally
with the quantity
\begin{equation}
   E_l(\Omega) = D_l(i\Omega)/c\Lambda^2,
   \label{new_variable}
\end{equation}
where $\Omega\equiv\omega \exp[-\int^l dl' z(l')]$ represents the
unrenormalized frequency; substituting (\ref{new_variable}) into
(\ref{imaginary_rg_equation}) and accounting for
(\ref{useful_relation}) as well as the trivial renormalization of
$\hbar_l$, see (\ref{hbar_renormalization}), we arrive at
\begin{eqnarray}
   \partial_l E_l(\Omega)
   &\approx& 2E_l(\Omega)+K e^{(d+2\zeta -2 )l}
   \frac{E_l(\Omega)\Theta (l-l_c)}{1+E_l(\Omega)}\frac{1}
   {\int_{-\infty}^{+\infty}\frac{\eta d\Omega/c\Lambda^2}{1+E_l(\Omega)}}
   \nonumber\\
   &\equiv& 2E_l(\Omega)+\frac{E_l(\Omega)\Theta(l-l_c)}
   {1+E_l(\Omega)}B_l,
   \label{equation_for_e}
\end{eqnarray}
with the dimensionless constant $K$ given by
\begin{equation}
   K = 2\pi\epsilon\frac{\eta\Delta(0)}
   {(c\Lambda^2)^2\hbar}e^{-(d+2\zeta-4)l_c}
   \sim \frac{S_\eta}{\hbar} e^{-(d+2\zeta)l_c}
   \label{K}
\end{equation}
and
\begin{equation}
   B_l=
   \frac{K e^{(d+2\zeta-2)l}}
   {\int_{-\infty}^{+\infty}\frac{\eta d\Omega/c\Lambda^2}
   {1+E_l(\Omega)}}
   \sim \frac{U(L)}{\int_{-\infty}^{+\infty}\frac{\hbar d\Omega}
   {1+E_l(\Omega)}}.
   \label{B}
\end{equation}
For $l<l_c$ the renormalization of the dynamics due to disorder
can be neglected, as properly expressed by the step function
$\Theta(l-l_c)$ in (\ref{equation_for_e}). The quantity $B_l
\propto 1/\Gamma_l$ governs the fluctuation-induced smearing of
the cusp in the force-force correlator. The right-hand side of
(\ref{equation_for_e}) behaves differently for $E_l(\Omega)\ll 1$
and $E_l(\Omega)\gg 1$; using the approximation
\begin{equation}
   \frac{E_l(\Omega)}{1+E_l(\Omega)}
   \approx {\rm min} \left(E_l(\Omega),1\right)
   \label{approx_E}
\end{equation}
considerably simplifies the analysis but does not change the
result qualitatively. We rewrite (\ref{equation_for_e}) in the
corresponding ($l$-dependent) frequency regions in the form
\begin{eqnarray}
   {\partial_l E_l(\Omega)}
   &\approx& 2E_l(\Omega)+E_l(\Omega) B_l, \quad
   E_l (\Omega) <1,
   \label{equation_for_<1} \\
   {\partial_l E_l(\Omega)}
   &\approx& 2E_l (\Omega)+B_l, \quad E_l (\Omega) > 1.
\label{equation_for_>1}
\end{eqnarray}
As is readily seen from (\ref{equation_for_<1}) and
(\ref{equation_for_>1}) the function $E_l(\Omega)$ is an
increasing function of $l$ for all $\Omega$ and
$E_l(\Omega)\rightarrow\infty$ as $l\rightarrow\infty$. In
addition, if $E_0(\Omega )$ is a monotonically increasing function
of $\Omega$ then $E_l(\Omega)$ remains monotonic in $\Omega$ for
any $l$. Let us define the frequency $\tilde{\Omega}_l$ and the
scale $\tilde{l}_\Omega$ which solve the equation
\begin{equation}
   E_l(\Omega) = 1;
   \label{tO}
\end{equation}
the function $\tilde{\Omega}_l$ starts at $l=0$ with a finite
value $\tilde{\Omega}_0$ and decreases with increasing $l$. For
any $l>0$ we distinguish between the three regions $0<\Omega <
{\tilde{\Omega}}_{l}$, $\tilde{\Omega}_{l}< \Omega <
{\tilde{\Omega}}_0$, and ${\tilde{\Omega}}_{0}<\Omega $, see Fig.\
\ref{fig:Omega_l}.

In each of these regions the function $E_l(\Omega)$ can be found
explicitly in terms of $E_0(\Omega)$ and the yet-to-be-determined
function $B_l$,
\begin{eqnarray}
   E_{l}(\Omega ) &=& E_{0}(\Omega)
   \exp\biggl[2l+\int\limits_{0}^{l}
   \thinspace dl^{\prime}B_{l^{\prime}}\biggr],
   \quad 0<\Omega < {\tilde{\Omega}}_{l},
   \label{reg1} \\
   E_{l}(\Omega) &=& e^{2(l-{\tilde l}(\Omega ))}
   \biggl[1+\int\limits_{{{\tilde l}(\Omega )}}^{l} dl^{\prime}
   e^{-2\left(l^{\prime}-{\tilde l}(\Omega )\right)}
   B_{l^{\prime}}\biggr],
   \quad {\tilde{\Omega}}_{l}< \Omega < {\tilde{\Omega}}_{0},
   \label{reg2} \\
   E_{l}(\Omega) &=& E_{0}(\Omega)
   e^{2l}+e^{2l}\int\limits_{0}^{l} \thinspace dl^{\prime}
   e^{-2l^{\prime}}B_{l^{\prime}}, \quad
   {\tilde{\Omega}}_{0}<\Omega.
\end{eqnarray}
Note that we integrate (\ref{equation_for_<1}) and
(\ref{equation_for_>1}) subsequently in the first two regions
$0<l<\tilde{l}_\Omega$ and $\tilde{l}_\Omega<l$, while for $\Omega
> {\tilde{\Omega}}_{0}$ equation (\ref{equation_for_>1}) applies
for all values of $l$. Correspondingly, the integral
$\int_{-\infty}^{+\infty} \thinspace d\Omega /{\left
[1+E_{l}(\Omega )\right ]}$ that determines $B_l$, see (\ref{B}),
can be written as a sum of three terms. We will show below that
$B_{l}$ increases exponentially, implying that the boundary
${\tilde\Omega}_{l}$ is exponentially small and the first integral
extending over the interval $0 < \Omega < \tilde{\Omega}_l$ can be
neglected, hence
\begin{eqnarray}
   \frac{1}{2}\int\limits_{-\infty}^{+\infty}\thinspace
   d\Omega \frac{1}{1+E_{l}(\Omega)}
   &\approx&\int\limits_{{\tilde\Omega}_{l}}^{{\tilde\Omega}_{0}}
   \frac{d\Omega}{1+e^{2(l-{\tilde l}(\Omega ))}
   \biggl[1+\int\limits_{{{\tilde l}(\Omega )}}^{l}dl^{\prime}
   e^{-2\left(l^{\prime}-{\tilde l}(\Omega)\right)}
   B_{l^{\prime}}\biggr]}
   +\int\limits_{{\tilde\Omega}_{0}}^{+\infty}
   \frac{d\Omega} {1+E_{0}(\Omega)e^{2l}
   + e^{2l}\int\limits_{0}^{l} \thinspace
   dl^{\prime} e^{-2l^{\prime}}B_{l^{\prime}}}\\
   &\approx&
   \int\limits_{0}^{{\tilde\Omega}_{0}}\thinspace \frac{d\Omega}
   {1+G_{l}}+\int\limits_{{\tilde\Omega}_{0}}^{\infty}\thinspace
   \frac{d\Omega}{1+E_{0}(\Omega ) e^{2l}+G_{l}}.
   \label{two_integrals}
\end{eqnarray}
In the last equation we have approximated ${\tilde\Omega}_{l}
\approx 0$ and ${{\tilde l}(\Omega )}\approx 0$ and have
introduced the expression
\begin{equation}
   G_l\equiv e^{2l}\int\limits_{0}^{l}\thinspace dl^{\prime}
   B_{l^{\prime}} e^{-2l^{\prime}}\approx
   \frac{B_l}{{d\ln B_l}/{dl} - 2};
   \label{G_eqn}
\end{equation}
the last approximation applies if $B_l$ increases faster than
$e^{2l}$. Substituting $E_{0}(\Omega ) = \left (\eta\Omega + \rho
\Omega^2\right )/c\Lambda^2$ (see (\ref{initial_condition_1}) and
(\ref{new_variable})) into (\ref{two_integrals}) we see that the
second integral on the right-hand side of (\ref{two_integrals})
always dominates and we conclude that it is the
{\it high-frequency} behavior of $D_0(\omega)$ that controls
the large length scale renormalization of the dynamics.

Using the definition of $B_l$, see (\ref{B}) and evaluating the
integrals over $\Omega$ explicitly we obtain an implicit equation
for $B_{l}$ valid at large length scales
\begin{equation}
   B_l\sim \frac{ K e^{(d+2\zeta-2)l}}
   {{\rm min}\left[e^{-2l}
   \ln\displaystyle{\frac{e^{2l}\eta^2/\rho}
   {G_l c\Lambda^2}},e^{-l}\frac{\eta/\sqrt{\rho}}
   {\sqrt{G_{l}c\Lambda^2}}\right]},
   \label{B(l)}
\end{equation}
where we have ignored multiplicative factors of order unity. The
first term in the denominator applies for small $\rho$ at
intermediate length scales when $G_l$ is not too large, while the
second is relevant if the inertia is substantial ($\rho > \eta^2
e^{2l}/c\Lambda^2 G_l$) and for asymptotically large scales for
{\it any} non-zero $\rho$. Given that the approximation in
(\ref{G_eqn}) is valid, as is the case in the regimes of interest,
we have a non-linear differential equation for $B_l$; in its
simplest approximation with $G_l \approx B_l$ this reduces to an
algebraic equation which is readily solved. Using the definition
of $K$, Eq.\ (\ref{K}) and the expressions $L_{c} = \Lambda^{-1}
e^{l_c}$, $U_{c}\sim c(\xi^2/L_c^2)L_c^d$, $S_{\eta}\sim \eta
\xi^2 L_c^d$, and $S_{\rho}^2 \sim \rho \xi^2 U_c L_c^d$, we find
the result
\begin{equation}
   B_{l}\sim \frac{\rho}{\eta} \frac{c \Lambda^2}{\eta} K^2
   \big[e^{(d+2\zeta-1)l}\big]^2
   \sim
   \bigg[\frac{S_{\rho}}{\hbar} e^{(d+2\zeta-1)(l-l_c)}\bigg]^{2}
   \label{B_inertial}
\end{equation}
in the {\it inertial case}. The behavior is rather more
complicated for the dissipative case: at intermediate length
scales we have
\begin{equation}
   B_l \sim \frac{K\,e^{(d+2\zeta)l}}{\ln[(\eta^2/
   {\rho c \Lambda^2 K) e^{-(d+2\zeta-2)l}}]}
   \sim
   \frac{S_{\eta}}{\hbar}\frac{e^{(d+2\zeta)(l-l_c)}}
   {\ln[(S_\eta\hbar/S_\rho^2)
   e^{-(d+2\zeta-2)(l-l_c)}]}
   \label{B_diss}
\end{equation}
which increases slightly {\it faster} with length scale than in
the absence of the logarithmic factor. The crossover between the
dissipative and massive results appears at
\begin{equation}
   B_{l_I} \sim K \frac{\eta^2}{\rho c \Lambda^2} e^{2l_I}
   \sim
   \frac{S^2_{\eta}}{S_\rho^2}e^{2(l_I-l_c)};
   \label{B_I}
\end{equation}
comparing with (\ref{B_diss}) this translates into the length
scale
\begin{equation}
   L_I \sim L_c
   \left(\frac{S_\eta \hbar}{S^2_\rho}\right)^{1/(d-2+2\zeta)},
   \label{L_I}
\end{equation}
with a corresponding energy scale
\begin{equation}
   U_I \sim U_c {\left(\frac{L_I}{L_c} \right)}^{(d+2\zeta -2)}
   \sim U_c \frac{S_\eta\hbar}{S^2_\rho} \sim
   \frac{\hbar\eta}{\rho}.
   \label{U_I}
\end{equation}
For $L > L_I$ the behavior of $B_l$ is always dominated by the
inertia and the result (\ref{B_inertial}) takes over.

In the end we see that the coefficient $B_l \propto 1/\Gamma_l$
describing the fluctuations rounding the cusp in the correlator
$\Delta_l$ increases dramatically with increasing scale $l$ (with
a correspondingly decreasing $\Gamma_l$). Substituting $E_l
\rightarrow E_0 e^{2l} + G_l \rightarrow B_l + (\Omega/
\Omega_c)^\alpha$ in the integral of (\ref{B}), we see that the
integration is squeezed to the high energy side where it is
ultimately cutoff by the inertial term ($\alpha = 2$) or a more
general cutoff $(\Omega/\Omega_c)^\alpha$. Hence the remaining
high-frequency fluctuations measured with respect to the typical
barriers $U(L)$ at this scale determine the smoothing coefficient
$\Gamma_l$. Technically, the exponent in the non-Arrhenius type
law (\ref{B_inertial}) then appears via solution of the implicit
equation for $B_l$ with the result $B_l \propto 1/\hbar^\alpha$.
For the extreme case with a linear spectrum sharply cut at the
frequency $\Omega_c$, $D_0 (i\omega< i\omega_c) = \eta \omega$ and
$D_0 (i\omega> i\omega_c) = \infty$, a similar calculation
provides an action that is {\it exponentially} (rather than
power-law) enhanced in the limit of small forces $f$.

Substituting the expressions for $B_{l}$ into the equation for
$E_{l}$ we can find how the dynamic spectrum $D_l(i\omega)$ is
renormalized. At low frequencies
\begin{equation}
   \partial_l D_{l}(i\omega\rightarrow i0)
   \sim [2+B(l)]D_{l}(i\omega\rightarrow i0),
   \label{small_frequency_equation}
\end{equation}
and hence $D_{l}(i\omega\rightarrow 0 )\propto \exp\left
[2l+\int_{0}^{l}dl^{\prime}B_{l^{\prime}}\right]$; consequently
the renormalized viscosity $\eta_f$ on the scale $l_f$ is
\begin{equation}
   \eta_{l_f}\approx
   \eta\exp\left[ 2l_f
   +\int_{0}^{l_f}dl^{\prime}B_{l^{\prime}}\right]\
   \label{eta_f}
\end{equation}
(note that if we had used the conventional normalization in the
action, adjusting the dynamic exponent $z(l)$ to keep $\eta_l$
fixed, we would have obtained the same results for physical
quantities but the renormalization would have gone into
$z(l)=2+B_l$ rather than the friction coefficient $\eta_l$).
\begin{figure}
\centerline{\epsfxsize = 7.0cm \epsfbox{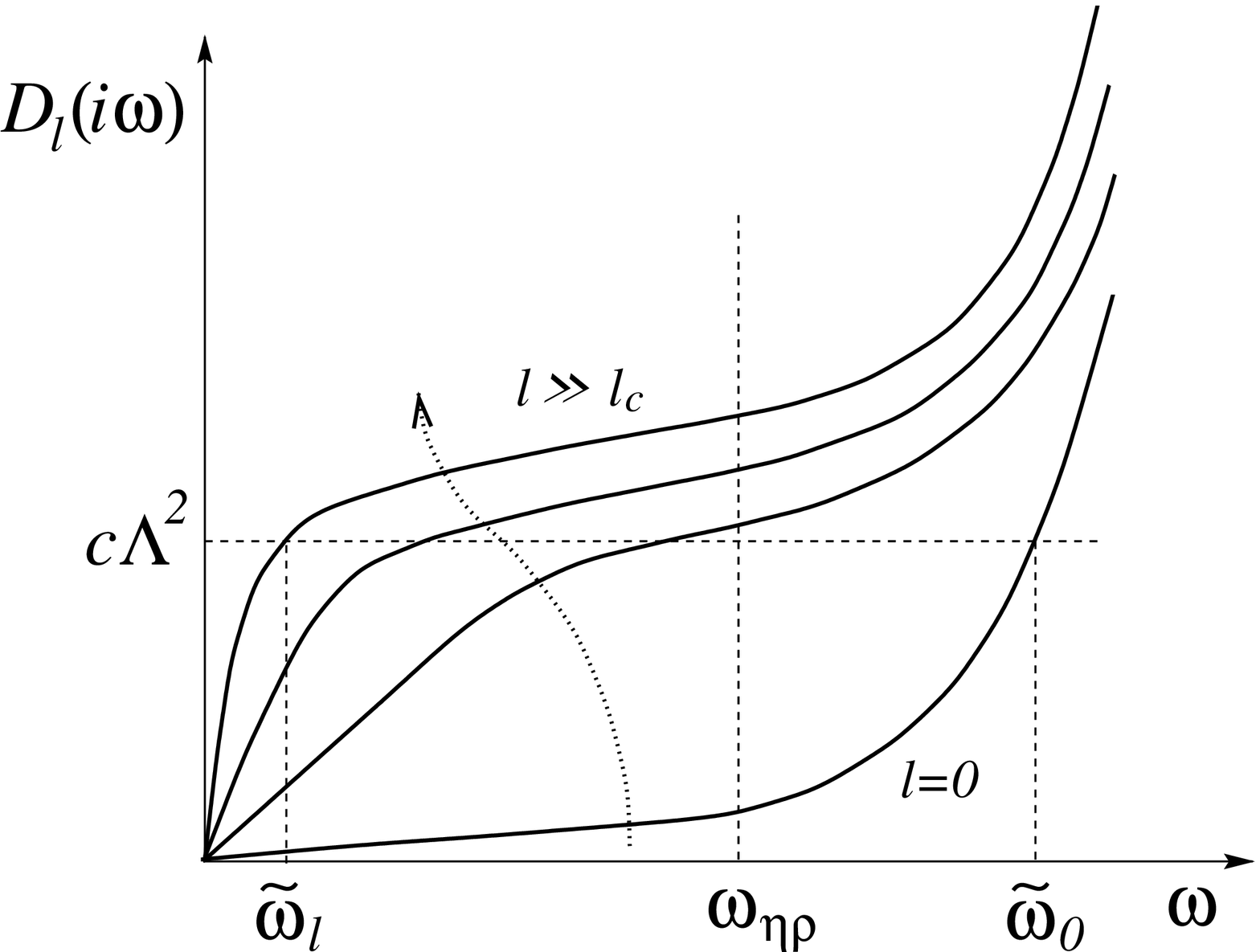}}
\vspace{0.3truecm} \caption{Schematic renormalization of the
dynamic spectrum $D_{l}(i\omega)$ for a dissipative/massive
initial dynamics, $D_{0}(i\omega ) = \eta |\omega | +
\rho\omega^2$, with a crossover at $\omega_{\eta\rho} =
\eta/\rho$. The spectrum is renormalized differently on small and
large frequencies: in the region $D_{l}(i\omega )\ll c\Lambda^2$
the spectrum is double--exponentially enhanced and grows as
$\eta|\omega| \exp[{\rm const.} e^{\beta l}]$, with $\beta
> 0$ an exponent depending on the specific dynamics. For
$D_{l}(i\omega )\gg c\Lambda^{2}$ the renormalization takes the
different form $D_l (i\omega) \approx \eta\omega e^{(2-z)l} +
\rho\omega^2 e^{(2-2z)l} + F_l$, with $F_l$ independent of the
frequency $\omega$.} \label{renormalization_of_dynamic_spectrum}
\end{figure}

Going back to the dynamical equation (\ref{equation_for_e}), we
see that $B_l$ not only determines the low-frequency part but the
entire function $D_l(i\omega)$ (we assume $z=0$ and identify
$\Omega$ with $\omega$). A schematic renormalization of the
dynamic spectrum $D_l (i\omega)$ is shown in
Fig.~\ref{renormalization_of_dynamic_spectrum}: the low-frequency
part at $\omega < \tilde{\omega}_l\sim (c\Lambda^2/\eta) \exp( -
B_l)$ where $D_l < c\Lambda^2$ remains linear, $D_l(i\omega) \sim
\eta_l \omega$, but is boosted exponentially with $\eta_l \sim
\eta \exp(B_l)$ (with $B_l$ itself growing exponentially in $l$).
At frequencies above $\tilde{\omega}_l$ the upward renormalization
is reduced and the response $D_l(i\omega)$ develops a flat
intermediate regime. Finally, at high frequencies $\omega >
\tilde{\omega}_0$ the renormalization remains small, while the
shape of the dynamical response again reflects the form of the
original bare dynamics $D_0$ with an additional shift $\propto
c\Lambda^2 B_l$. It is this high-frequency part of $D_l$ that
dominates the important renormalization of $\eta_l$ at low
frequencies. We attribute the strongly renormalized low-frequency
part $D_l < c \Lambda^2$ to those degrees of freedom of the
manifold describing its inter-valley motion, while the remaining
modes at intermediate and high frequencies describe its
intra-valley motion smoothing the disorder landscape. In the next
section we will discuss the meaning, significance, and problematic
aspects of the unusual dynamical renormalization scheme uncovered
above.

\subsection{Results: Quantum Creep} \label{results_for_velocity}

The physical quantities of primary interest can be obtained from
the analysis of the previous section. In particular, using the
relation (\ref{relation}) we find the creep velocity at low forces
in the inertial case,
\begin{equation}
   v \propto \exp\biggl\{-\biggl[\frac{S_{\rho}}{\hbar}
   {\biggl(\frac{f_{c}}{f}\biggr)}^{(d+2\zeta-1)/(2-\zeta)}\biggl]^2
   \biggr\},
   \label{inertial_answer}
\end{equation}
with an unknown multiplicative coefficient of order unity
incorporated into $S_\rho$. In the dissipative case there is a
characteristic crossover force
\begin{equation}
   f_I \sim f_c
   \left(\frac{\rho U_c}{\hbar\eta}\right)^{2/(d-2+2\zeta)}
   \label{f_I}
\end{equation}
that separates two distinct regimes: For intermediate forces $ f_I
\ll f\ll f_c$ we have
\begin{equation}
   v \propto \exp \biggl[-\frac{S_{\eta}}{\hbar}
   {\biggl(\frac{f_{c}}{f}\biggr)}^{(d+2\zeta)/(2-\zeta)}
   \frac{1}{\ln(f/f_I)}\biggr]
   \label{dissipative_answer}
\end{equation}
with an unknown constant coefficient incorporated into $S_\eta$.
But at asymptotically low forces $f\ll f_I$ the inertial term in
the action dominates and the behavior of the velocity crosses over
to the inertial result (\ref{inertial_answer}). For the
dissipative case, the result is similar to that obtained using
scaling arguments in section~\ref{dissipative_dynamics} in the
{\it intermediate} regime, but the logarithmic factor causes the
velocity to decrease slightly faster with decreasing force than
anticipated. By contrast, in the inertial case the creep velocity
is much smaller that anticipated: it can be written in the form $
v \propto \exp \bigl[-{(S_{\rm mass}(f)/\hbar)}^2\bigr]$, with
$S_{\rm mass}(f)$ the characteristic action obtained in
section~\ref{inertial_dynamics} using scaling arguments.  At
asymptotically low forces, this result is also valid for the more
general model including a dissipative dynamics at low frequencies
and an inertial dynamics at high frequencies.

\subsection{Results: Classical Creep and Crossover}
\label{cl_cr}

At high temperatures the coefficient $\Gamma_l =
C^{>}_{l}(\Lambda,t = 0)$ is independent of the dynamics and the
evolution of the spectrum $D_l(i\omega )$ does not feed back into
$\Gamma_l$. Equation (\ref{small_frequency_equation}), written in
terms of the bare temperature, then takes the form
\begin{equation}
   {\partial_l} D_{l}(i\omega\rightarrow 0)
   \sim
   \frac{U(L)}{T} D_l(i\omega\rightarrow 0).
   \label{classical_small_frequencies}
\end{equation}
Using the relations (\ref{relation}), $U(L) \sim U_c
(L/L_c)^{(d+2\zeta-2)}$, $L \sim L_c (f/f_c)^{1/(2-\zeta)}$, we
obtain the classical creep law $v(F) \sim \exp[-U(f)/T]$ with
$U(f) \sim U_c {\left (f_c/f\right )}^{(d+2\zeta-2)/(2-\zeta)}$,
see Ref.\ \onlinecite{Chauve}. The exponent $\theta=d+2\zeta-2$
that determines the scaling of $U(L)$ is simply the scaling
dimensionality of the temperature, see
(\ref{temperature_renormalization}), in terms of which
(\ref{classical_small_frequencies}) can be written with a
prefactor $U(L)/T \sim U_c/T_l \sim e^{(d+2\zeta-2)(l-l_c)}
U_c/T$.

One can find the crossover temperature from classical to quantum
creep by comparing the exponents in (\ref{dissipative_answer}) and
(\ref{inertial_answer}) with $U(f)/T$. For the inertial case,
\begin{equation}
   T_{\rm cr}\propto f^{(d+2\zeta)/(2-\zeta )},
   \label{T-cross}
\end{equation}
which is different from the naive result obtained via simple
scaling arguments, see section~\ref{scaling_arguments}. For the
dissipative case, the crossover temperature depends on the regime.
For intermediate forces, it is the same as that given by the naive
scaling arguments (up to a logarithmic factor), but for $f\ll f_I$
the crossover temperature is again given by the result
(\ref{T-cross}) for the inertial case.

\subsection{Interpretation}

As mentioned earlier, a dependence on $\hbar$ of the form
(\ref{inertial_answer}) might be expected if the dynamics were
dominated by atypical barriers. A simple example is given by a
classical particle diffusing in a short-range correlated random
potential. At positive temperatures there is a linear response to
an external force with the inverse mobility of the particle
proportional to $\int\thinspace dU {\cal P}(U)\exp(U/T)$, where
${\cal P}(U)dU$ is the probability density of barriers $U$. For
the case of a Gaussian distribution of the random potential, we
see that $v \propto f\exp(-{\rm const.}/T^{2})$, resulting in a
non-Arrhenius temperature dependence.

Analogous effects can occur in quantum transport: Consider a
quantum particle of mass $m$ and with a dissipative coefficient
$\eta_l$, tunneling through a succession of barriers of random
heights $U$ but, for simplicity, uniform width $a$. The inverse
mobility of such a particle can be written as
\begin{equation}
   \frac{1}{\mu} \propto
   \int\thinspace dU {\cal P}(U) \exp
   \biggl(\frac{\eta a^2}{\hbar}+\frac{\sqrt{mU}a}{\hbar}\biggr).
   \label{example_tunneling}
\end{equation}
The exponent contains a sum of two actions, the first due to
dissipation and the second describing the inertial response. One
can see that even if the dissipation is strong, in the limit
$\hbar\rightarrow 0$ the inertial effects can dominate: the
integral in (\ref{example_tunneling}) is calculated using the
method of steepest descent and since the massive contribution to
the action is proportional to $\sqrt{U}$, it will contribute a
term larger than the dissipative one. In particular, with a
Gaussian distribution of barriers of the form ${\cal
P}(U)\sim\exp[-(U/U_0)^2]$ this leads to $\mu \propto
\exp[-{({\sqrt{mU_0}a}/{\hbar})}^{4/3}]$ and we obtain a similar
non-trivial dependence on $\hbar$ as found above in
(\ref{inertial_answer}) for the creeping elastic manifolds of
interest here.

In this simple single-particle example it is easy to understand
what is going on: Because the particle must tunnel through a
succession of barriers, the dynamics is dominated by the largest
ones as long as there is a sufficiently long tail to the barrier
distribution; the smaller the quantum fluctuations, the larger the
barriers that dominate. For the probability distribution function
${\cal P} (U)$ chosen above, $U\simeq \left (\sqrt{m}a
U_0^2/\hbar\right )^{2/3} \rightarrow\infty$ for $\hbar\rightarrow
0$. The form of the tail of the distribution of barriers thus
dominates the mobility and gives rise to the unusual dependence on
$\hbar$. The fact that for small $\hbar$ the dominant barriers are
high implies that the characteristic time scale for tunneling
through the barrier is short (in the above case the time $\tau$ is
given through $\tau\simeq a\sqrt{m/U}$); this is what causes the
long-time behavior as manifested in the mobility to be dominated
by the inertia rather than by the dissipative response. In the
case of interest here, the elastic manifold, the barriers that
must be surmounted depend on the driving force --- the lower the
force, the higher these barriers are. If one assumes that the
barriers relevant for tunneling have an unbounded distribution
similar to the toy model studied above, it is not surprising that
it is the inertial dynamics that dominates the limit of low
forces.

However, the reason for the anomalous dependence on $\hbar$ is
more subtle for the elastic manifold with its many degrees of
freedom. As can be seen from the analysis in the previous section,
the anomalous dependence can be traced back to the dependence of
the quantum fluctuations on one length scale, as parametrized by
$\Gamma_l$, on the random pinning at smaller scales. Very crudely,
this might be interpreted as leading to an increase of the {\it
effective mass density} with length scale caused by the motion of
smaller scale sections of the manifold implicit in the tunneling
motion of a segment of size $L_f$. Determining whether or not this
is a reasonable interpretation must wait for a better
understanding of the physics underlying the RG results and whether
these are valid.

\section{Validity of RG Results}
\label{validity}

In both the previous and the present work on classical and quantum
creep of elastic manifolds the validity of the approximations that
underly the RG formulation have not been carefully examined. In
unpublished work \cite{RBF} one of the problematic aspects, the
possible effects of tails in the distribution of local effective
friction coefficients, has been investigated. Here, we briefly
summarize the potential problems that this suggests as well as
more basic ones that have not, to our knowledge, been raised
previously.

\subsection{Random Friction}
\label{random_friction}

One difficulty, analyzed in Ref.\ \onlinecite{RBF}, is already
apparent in the toy model of a single particle in a random
potential: the broad distribution of times to go through or over
barriers. In particular, as discussed in section
\ref{results_for_velocity} above, for a single particle the
mobility is dominated not by the typical or even the average rate
for overcoming the local barriers, but by the {\it average time}
to overcome them; and the average time is dominated by anomalously
large barriers. As this problem already arises in the classical
case both in the toy model and for elastic manifolds, we restrict
our discussion to the simpler classical limit.

The main idea of the RG is to derive equations which relate the
renormalized parameters of the field theory to the bare ones. Very
often the parametric space of the bare and renormalized theories
are identical. In other words, if the bare theory is described by
the parameter set $\alpha_{0}^{(1)}, \alpha_{0}^{(2)}\dots$ the
renormalized theory will be described by the same set of variables
$\alpha_{l}^{(1)}, \alpha_{l}^{(2)}\dots$. An example of this kind
of RG is the $\phi^4$ theory to one-loop order. It could be,
however, that under the RG flow additional variables
$\beta_{l}^{(1)}, \beta_{l}^{(2)}\dots$ are generated even if
their bare values are zero. These variables might be strongly
relevant and feed back to the original set of parameters. An
example of such a behavior, which we could handle successfully,
has already been considered above: in order to obtain sensible
results we had to introduce a function $D_l(i\omega)$ describing
the dynamics on {\it all} frequencies. In this paragraph we will
show that another set of dangerous variables is generated under
the RG flow --- these variables describe the probability
distribution function of waiting times. In the RG scheme
considered above the most crucial quantity is the renormalized
viscosity $\eta_l$ which is proportional to the waiting time at
the scale $l$. We will show that the randomness due to the
point-like disorder will produce a random and spatially
inhomogeneous distribution of frictions which appears to be very
broad and hence cannot be properly described by its first moment
$\eta_l$ alone.

Let us then consider randomness in the local effective friction
coefficients from the beginning and assume that the friction
$\eta$ is a function $\eta = \eta [u({\bf z}),{\bf z}]$ of both
the displacement $u({\bf z})$ and the internal coordinate ${\bf
z}$ of the manifold. The local $\eta$ has the natural
interpretation as a characteristic time to overcome barriers
involving the smaller length scale deformations that have already
been integrated out. The simplest case to consider is a random
potential that is periodic in $u$ with a locally random phase
shift; such a model is applicable to charge density waves (CDWs).
Because of the periodicity, for CDWs we can expand the function
$\eta(u,{\bf z})$ into the Fourier series and there will be a
component that is independent of $u$ and only depends on ${\bf
z}$; we consider the effects of such randomness here.

We assume that $\eta({\bf z})$ is a random short-range correlated
variable with cumulants $\eta^{(n)}$; e.g., the first three take
the form $\overline{\eta({\bf z})} = \eta^{(1)} \equiv \eta$,
$\overline{\eta({\bf z})\eta({\bf z'})} - \overline{\eta({\bf
z})}\thinspace\overline{\eta({\bf z}^{\prime})} = \eta^{(2)}\delta
({\bf z} - {\bf z}^{\prime})$, and
\begin{eqnarray}
   \overline{\eta ({\bf z}) \eta ({\bf z}^{\prime})
   \eta ({\bf z}^{\prime\prime})}  -
   \overline{\eta ({\bf z}) \eta ({\bf z}^{\prime})}
   \thinspace\thinspace\thinspace
   \overline{\eta ({\bf z}^{\prime\prime})} -
   \overline{\eta ({\bf z}^{\prime\prime}) \eta ({\bf z}^{\prime})}
   \thinspace\thinspace\thinspace
   \overline{\eta ({\bf z})} -
   \overline{\eta ({\bf z}^{\prime\prime}) \eta ({\bf z})}
   \thinspace\thinspace\thinspace
   \overline{\eta ({\bf z}^{\prime})}
   + 2{\overline{\eta ({\bf z})}}^3\nonumber\\
   =
   2\eta^{(3)}
   \left[\delta \left ({\bf z} - {\bf z}^{\prime}\right ) \delta
   \left({\bf z}-{\bf z}^{\prime\prime}\right )
   +\delta\left({\bf z}^{\prime}-{\bf z}^{\prime\prime}\right)
   \delta\left({\bf z}^{\prime}-{\bf z}\right )
   +\delta\left({\bf z}^{\prime\prime}-{\bf z}\right)
   \delta\left({\bf z}^{\prime\prime}-{\bf
   z}^{\prime}\right)\right].
   \label{cumulant_3}
\end{eqnarray}
Note that we define cumulants $\eta^{(n)}$ up to a factor $n!$.
After averaging over the randomness the classical MSR-action
(\ref{MSR_action}) will have additional terms of the form
\begin{eqnarray}
   {A_{\rm rand}} =
   \sum\limits_{n\ge 2}{A}^{(n)} =
   \sum\limits_{n\ge 2}{(-1)}^{n}\eta^{(n)}
   \int\thinspace d^{d}z dt_{1}\dots dt_{n}\thinspace
   {\dot u}({\bf z},t_1)iy({\bf z}, t_1)\dots
   {\dot u}({\bf z},t_n)iy({\bf z}, t_n) .
   \label{dangerous}
\end{eqnarray}
When deriving the RG equations with an action that includes terms
of the form $A_{\rm rand}$, it is necessary to find the average
over fast modes of terms containing products of two perturbations,
in particular, terms of the form
\begin{equation}
   \delta A = \frac{1}{2}\left \langle A_{\rm rand}
   \int\thinspace d^{d}z\tau_1 d\tau_2\thinspace\Delta_l
   \left[u({\bf z},\tau_1)-u({\bf z},\tau_2)\right]
   iy({\bf z},\tau_1) iy({\bf z},\tau_2)\right\rangle_{>},
   \label{cross_average}
\end{equation}
with $\langle\dots\rangle_{>}$ the standard RG average over fast
modes. Making use of (\ref{dangerous}), the average
(\ref{cross_average}) can be written as a sum of terms involving
the cumulants $\eta_l^{(n)}$ (see appendix~\ref{appendix2} for
details; here, we summarize the main ideas of the calculation).
The term of order $n$ in (\ref{cross_average}) then generates $2n$
terms proportional to $\delta {A}^{(n)}_1$,
\begin{eqnarray}
   \delta {A}^{(n)}_1 =
   {(-1)}^{n}\eta_l^{(n)}\Delta^{\prime\prime}_{l}(0)
   \int\thinspace d^{d}z dt_{1}\dots dt_{n}\thinspace
   {\dot u}({\bf z},t_1)iy({\bf z}, t_1)\dots
   {\dot u}({\bf z},t_n)iy({\bf z}, t_n)
\end{eqnarray}
which feed back to the term (\ref{dangerous}). In addition, there
are $4n^2-4n$ terms proportional to $\delta {A}^{(n)}_2$,
\begin{eqnarray}
   \delta {A}^{(n)}_2 =
   {(-1)}^{n}\eta_l^{(n)} \int\thinspace d^{d}z dt_{1} \dots dt_{n}
   \thinspace
   {\dot u}({\bf z},t_1)iy({\bf z},t_1)\dots
   \thinspace\nonumber\\
   {\dot u}({\bf z},t_n)iy({\bf z},t_n)
   \Delta^{\prime\prime}_{l}
   \left[u({\bf z},t_1)-u({\bf z},t_2)\right] .
\label{super_dangerous}
\end{eqnarray}
These renormalizations involve the behavior of the correlator
$\Delta_{l}(u)$ at the origin whose growth at long length scales
is crucial for the renormalization group analysis, see the above
discussion. We can keep track of the most dangerous terms by
substituting $\Delta^{\prime\prime}_{l}(u)$ by $\Delta^{\prime
\prime}_l(0)$ in (\ref{super_dangerous}). The renormalization of
the cumulants $\eta_l^{(n)}$ from the above process can be written
in the form
\begin{equation}
   \partial_l \eta_l^{(n)} \propto -\Delta^{\prime\prime}_{l}(0)
   \left(2n^2-n\right)\eta_l^{(n)} ;
   \label{dangerous_rg}
\end{equation}
furthermore, there are non-linear terms that create higher moments
from lower moments, see appendix~\ref{appendix2}. As
$-\Delta^{\prime\prime}_l(0)$ grows exponentially with $l$, i.e.,
as a power of the length scale, the cumulants $\eta_l^{(n)}$ grow
very rapidly. The $n^{2}$ coefficient and the positivity of
$-\Delta^{\prime\prime}_{l}(0)\propto {1}/T_l\propto e^{\theta l}$
in Eq.~(\ref{dangerous_rg}) imply
that the high order moments grow so fast that ratios of the form
$\eta_l^{(n)} /\eta_l^n$, which naively are expected to be
dimensionsless (in the RG sense), {\it themselves} grow
exponentially with increasing length scale. Indeed, the high order
moments increase so rapidly with $n$ that, if these results are
taken literally, the distribution of $\eta_l({\bf z})$ has such a
long tail that it is not uniquely determined by its moments ---
and it is certainly not well characterized by its mean
$\overline{\eta_l ({\bf z})} = \eta_l$. Note that a random
friction $\eta_l({\bf z})$ is not dangerous near the
zero-temperature depinning transition ($f - f_c \ll f_c$) as in
this case $\Delta^{\prime\prime}_{l}(0)$ should be substituted by
$\Delta^{\prime\prime}_{l}(0+)>0$ and Eq.~(\ref{dangerous_rg})
suggests that $\eta_l^{(n)}$ renormalizes to zero for any $n$
(although it will actually be stabilized at a small value of the
order of an $n$-dependent power of $\epsilon$ because of other
terms).

The analysis in appendix~\ref{appendix2} shows that even if
initially the friction is non-random, the disorder term alone will
generate the corrections to the second cumulant. The second
cumulant will then generate the third cumulant to the next loop
order, and so forth. As the cumulants grow extremely rapidly, the
RG flow becomes essentially uncontrollable. However note, that
there is still an approximation in the above analysis: we have
substituted the argument of the second derivative of the
disordered correlator $\Delta_l^{\prime\prime} (u)$ by zero in
Eq.~(\ref{super_dangerous}). In order to be accurate we have to
include the terms of the form
\begin{eqnarray}
   \delta {A}^{(n)}_2 =
   {(-1)}^{n}\eta_l^{(n)} \int\thinspace d^{d}z dt_{1} \dots dt_{n}
   \thinspace
   {\dot u}({\bf z},t_1)iy({\bf z},t_1)\dots
   \thinspace\nonumber\\
   {\dot u}({\bf z},t_n)iy({\bf z},t_n)
   \sum\limits_{i\ne j}^{n}
   F_{l}
   \left[u({\bf z},t_i)-u({\bf z},t_j)\right] ,
\label{super_dangerous_new}
\end{eqnarray}
cf.\ Eq.~(\ref{super_dangerous}), into the action, i.e., we have
to renormalize one more function $F_l (u)$. Under the RG
transformation this function will produce a new set functions, and
so forth --- it is presently unclear how all these variables can
be analyzed in a regular way.

One way that one might hope to make progress is to rewrite the
equations of motion so that instead of having to deal with a
random $\eta$ on the left hand side, one works with a random
mobility $\mu=1/\eta$ on the right hand side. This quantity, as it
is bounded from above by the fastest motion, is unlikely to have
troublesome long tails in its distribution. But the appearance of
a random mobility multiplying all the spatially random pinning
forces introduces additional technical complications into the
formalism and it is presently not clear how to handle them.
Nevertheless, there are a lot of constraints on the
renormalization, e.g., the {\it static} response will not be
modified by the randomness of either the pinning or the mobility
at any wavelength. Whether this is enough to make possible a fully
controlled analysis of at least the classical thermal creep regime
is an interesting challenge.

\subsection{Underlying Formulation} \label{RG-formulation}

It is possible that the apparent runaway of the distribution of
the local friction coefficients is indicative of a breakdown in
the basic scaling assumption that underlies this and earlier work:
If the barriers for motion scale with an exponent $\psi$ that is
larger than the exponent $\theta=d-2+2\zeta$ controlling the
scaling of fluctuations in minimal energies and the
renormalization of the inverse temperature, then the present
scheme where the dynamics is controlled by static properties, such
as the correlator $\Delta(u)$, cannot be valid. One then has to
take into account all the dangerous variables discussed above.

It is instructive to go back to the original formulation of the RG
expansion for the depinning in the absence of fluctuations
\cite{NF-CDW}, henceforth NF. In the derivation of the $\phi^4$
theory from the Ising model for conventional equilibrium phase
transitions, the starting point is an expansion around a mean
field theory and the actual ``field" $\phi({\bf x})$ used in the
RG formulation is closely related to the local effective field ---
applied plus exchange ---  acting on a spin rather than to the
spin itself. If the interactions are long but finite range, these
fields will be slowly varying in space and weakly fluctuating,
enabling a systematic expansion to be started.

NF focus on one segment ${\bf z}$ and use the linear combination
$u({\bf z},t)$ of the displacements of {\it other segments} that
determine the elastic force on ${\bf z}$ as the basic field, which
is hence intrinsically a coarse-grained quantity. The underlying
local displacements we will here call $w({\bf z},t)$. The segment
${\bf z}$ feels a linear restoring force proportional to $u({\bf
z},t)-w({\bf z},t)$, plus the applied driving force, plus a
quenched random pinning force that is a function of $w({\bf z})$,
and thermal noise. The vertices in the effective field theory are
given by correlations and responses of $w({\bf z},t)$ to the
time-dependent fields $u({\bf z},t)$. In particular, the
force-force correlator $\Delta (u)$ that plays an essential role
is related to the average
\begin{equation}
   \Upsilon \equiv \overline
   {\langle\big[w({\bf z},t)-w({\bf z},t')\big]^2\rangle}
   \label{Ups}
\end{equation}
over the random pinning forces and, at positive temperature,
thermal noise. In general, $\Upsilon$ is a {\it functional} of
$u({\bf z},\tau)$ over all times $\tau$. For zero temperature
depinning, the case of primary interest, the possible fields $u$
are limited to those that are non-decreasing in time. In this
case, it can be seen that the crucial parts of $\Upsilon$ (which
are sufficient to analyze the depinning critical behavior) depend
only on $u({\bf z},t)-u({\bf z},t')$ and the functional $\Upsilon$
simplifies to a function of this one variable: it is then of the
form assumed for $\Delta (u)$. In particular, if $u$ does not
change between $t$ and $t'$, $w$ will not change either unless the
$w$ at the earlier time was the cause of a jump out of a formerly
stable configuration into another; this jumping case can be
handled by putting in time delays into the definition of $u$ and,
beyond this, the local dynamics will be independent on the history
of $u$ prior to times $t$ and $t'$.

As soon as one considers a more general dynamics --- e.g., still
fluctuationless but with the applied force allowed to decrease
with time, non-monotonic stress transfer kernels, or thermal noise
--- the simplification of the functional $\Upsilon$ does not
occur. In general, it is then not clear how to proceed. In the
case of interest for the present paper, one could first assume, as
in all expansions about a mean field theory, that the fields are
slowly varying in space and time and weakly fluctuating about a
uniformly advancing solution which has a slow mean velocity $v$.
The local displacements will lag behind due to the pinning but
will be pulled ahead by the driving force: the balance of these
effects determines the velocity--force relation.  With thermal
fluctuations, the displacements will lag less than they would
otherwise because at low velocities they have the time to surmount
energy barriers. Consider now the effects of a time dependent
change in $u({\bf z},t)$ on $\overline{\langle w({\bf
z},t)\rangle}$. If this change is very slow, $w$ will follow
approximately adiabatically. But if the change is relatively fast
--- as it can be due to the fast motion of a neighboring segment
over a barrier --- how the local displacement $w$ will respond
will depend crucially on whether it has already been near a
surmountable barrier for some time, and thus is likely to have
already surmounted it, or has recently arrived near a barrier and
could thus be pushed over it by the change in $u$. Thus we expect
the responses and correlations of $w$ to depend on the whole prior
(and intervening) history of the fields $u$.

The basic issue that must be addressed to make further progress is
whether or not the essential information about the basic
activation processes can be subsumed in simplified functions, such
as the force-force correlator $\Delta(u-u')$, that appear in the
formulation used in the present paper (note, however, that as we
are performing an epsilon expansion, the central limit theorem is
likely to be helpful: as we go from one scale to another, we can
always assume that there is a large number of segments (or
effective intermediate scale segments) whose dynamics contributes
to the fields on larger scales). One possible route to proceed is
to start by considering the nature of the typical fields that will
arise from the thermally activated motion of large segments on the
dynamics at smaller scales and ask whether the local responses to
these are typically simple enough to be captured by the type of
approximations used here.

We leave further work on these points as a challenge for the
future. It is worth noting, however, that the physical origin of
the problems, the intrinsic history dependence of the phenomena of
interest, arise in many other non-equilibrium systems with many
degrees of freedom. Better understanding of them in this context
is thus likely to bear fruit more generally.

\subsection{Alternative Expansions}

Many of the difficulties encountered in this paper would be
lessened if the energy scale --- and hence the barriers --- did
not grow rapidly with length scale. One way to get around this
difficulty might be to consider an expansion about a different
limit, one in which not only the roughness exponent $\zeta$ is
small, but also the scaling exponent $\theta$ for the energy. In
$d$-dimensional manifolds with long range elastic interactions
that fall off as $1/r^{d+\alpha}$ with $0<\alpha<2$, the upper
critical dimension for both fluctuationless depinning and rough
manifolds in equilibrium is $d_c=2\alpha$ (i.e., $\zeta = 0$ at
$d=d_c$). The energy scaling exponent is $\theta=d-\alpha+2\zeta$
so that for $\alpha$ small, $\theta(d_c)=\alpha$ is also small.
One might then hope to attempt an expansion in $\alpha$ and
$\epsilon=d_c-d$ simultaneously. This would have the advantage
that, assuming the barrier scaling exponent $\psi$ is also small
for small $\alpha$ near $d_c$, the exponential dependence of the
velocity on the length dependent barriers would be relatively weak
and hence, perhaps, systematically controllable. We must also
leave this and other possible limits about which more controllable
expansions might be attempted for future research.

\subsection{Quasiclassical Langevin Equation}

In addition to potential problems with the renormalization group
formulation that are analogous to those discussed above for
classical creep, in order to obtain results for quantum creep we
have resorted to a quantum Langevin equation to approximate the
quantum dynamics. We would like to be able to investigate this
formalism more deeply and understand its limitations and regimes
of validity. It is instructive to first consider simpler problems.
\begin{figure}
\vspace{0.3truecm} \centerline{\epsfxsize = 7.0cm
\epsfbox{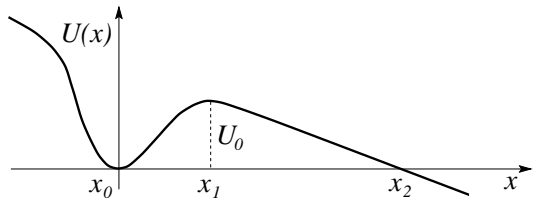}} \vspace{0.3truecm} \caption{Tunneling
of an overdamped particle out of a 1D potential well in the
presence of an external force $f$. The quantum mechanical answer
for the tunneling action is of the form $S\propto 1/{f^2}$,
$f\rightarrow 0$. The action obtained via the QLE is determined
only by the region $x_0 < x < x_1$. Since $x_1 - x_0 \ll x_2 - x_1$,
the QLE overestimates tunneling effects (i.e. gives smaller
action).} \label{QLE_illustration}
\end{figure}

The QLE can be used to analyze tunneling of a single particle if
the barrier separating the metastable state from the lower-energy
stable one has a single characteristic height and width. In this
case the tunneling action obtained from the QLE differs from the
true one only by a factor of order unity. In principle, the QLE
can lead to an overestimation of the tunneling rate. For example,
consider an overdamped particle (with viscosity $\eta$ )
in a 1D potential well $U(x)$ of depth $U_0$ with a
minimum at $x=0$ and $U(x)$ rapidly decaying for
$x\rightarrow\pm\infty$. In the presence of an external force $f$,
see Fig.~\ref{QLE_illustration}, the well becomes metastable and
the particle will leave. The quantum action $S$ can be estimated as
\begin{equation}
   S \simeq \eta {(x_2 - x_1 )}^2 \approx
\eta {\left (U_0/f\right )}^2 .
   \label{S_wkb}
\end{equation}
This action is dominated by a region of width $\sim 1/f$ where the
potential $U(x)$ is negligible but $fx$ is less than the depth of
the well: this yields $S\propto 1/{f^2}$ as $f\rightarrow 0$. If we
use the QLE approximation, the tunneling action would be much
smaller than the true WKB value in the limit $f\rightarrow 0$:
since the QLE is a {\it local} approximation, the only relevant
part of the potential would be that for $x<x_1$, with $x_1$ the
point where $U(x)-fx$ is maximal. In the QLE, once the random
force that obeys the quantum fluctuation--dissipation theorem
brings the particle to the point $x_1$, the particle can leave the
well and the region $x>x_1$ that dominated the WKB action does not
matter. Since in the limit $f\rightarrow 0$, $x_1 - x_0\ll x_2 -
x_1$, the tunneling action obtained within such an approach is
much smaller than the correct one, $S_{QLE}\simeq \eta {(x_1 -
x_0)}^2\ll S$; the QLE can thus drastically {\it overestimate}
tunneling rates.

We believe, however, that for the quantum creep problem considered
in this paper the QLE should be a reasonably accurate
approximation for the action whose exponential determines the
velocity (\ref{inertial_answer}): in the naive scaling arguments,
the effective potential wells that segments must tunnel out of
have only one characteristic width and height. Nevertheless, the
results we obtain from the QLE and RG analysis yield a much larger
effective action for the tunneling than the naive scaling
arguments. The source of this is unlikely to be due to the QLE
approximation as, by analogy with the single particle case above,
we would expect that, if anything, then the QLE would
underestimate the tunneling action. This argument is reinforced by
noting that the scaling analysis in
section~\ref{scaling_arguments} can be applied directly to a
system driven by a stochastic force obeying the quantum FDT for
which the QLE is exact: these would lead to the same results as
those obtained from the scaling arguments for the full quantum
dynamical problem, see also Ref.~\onlinecite{Eckern}.

\section{Conclusions}
\label{conclusion}

In this paper we have considered the motion of an elastic manifold
driven through a disordered medium at low temperatures where
quantum fluctuations are important. We have focused on the limit
of small driving forces ($f\ll f_c$) for which the average
velocity of the manifold is small and dominated by quantum
tunneling through barriers between locally stable configurations.
Using an at-least-partially controlled RG expansion, we find
that the resulting creep velocity is exponentially small in a
power of $1/f$. While for strong dissipation we find an
intermediate range of forces where the creep law agrees with the
result of simple scaling estimates (up to logarithmic corrections,
cf.\ (\ref{dissipative_action_scaling}) and
(\ref{dissipative_answer})), we find that the results of the RG
analysis are more complex. In particular, at asymptotically small
forces the creep velocity has the non-trivial form
(\ref{inertial_answer}), combining an underlying massive (or other
high-frequency) dynamics with an unexpected non-Arrhenius-like
dependence on $\hbar$. The structure of the renormalization group
flow is rather subtle and very different from that for classical
creep. In particular, because of the importance of both rapid
motion during a tunneling event and the slow overall motion, it is
necessary to consider the dynamics of the manifold at {\it all}
frequencies $\omega$. This is in contrast to the classical case
for which Chauve {\it et al.} \cite{Chauve} considered only the
low-frequency limit of the dynamics. In the quantum case such a
low-frequency approximation would lead to a spurious ''localization
transition'' with the average velocity of the manifold dropping to
zero at a small but finite value of the driving force. Although
future improvements based on the analysis of the complete quantum
mechanical action (\ref{quantum_action}) rather than the quantum
Langevin approximation might change some of the results presented
here, we believe that the main feature, the importance of the
whole spectrum of frequencies for the tunneling dynamics, will
persist.

In spite of the subtleties that appeared in the analysis we have
carried out there are further difficulties associated with both
the broad distribution of barriers and the underlying field
theoretic formulation of expansions about mean field theory.
Although we have not resolved all the difficulties, and thus are
not sure whether the present results are truly systematic, we have
noted the physical and formal sources of the problems in ways that
we hope will help direct future progress.

\acknowledgments

DSF thanks L.\ Radzihovsky and L.\ Balents for the collaboration
that led to the understanding of some of the subtleties in the RG
scheme discussed in the last section. We thank J.-P.\ Bouchaud and
D.R.\ Nelson for helpful discussions. This work was supported by
the National Science Foundation under grants DMR 9976621, DMR
9809363, and DMR 9714725. We thank the Swiss National Science Foundation for
financial support.

\begin{appendix}
\section{Derivation of the Real-Time Quantum Action}
\label{appendix}

In this appendix we discuss the derivation of the action
(\ref{quantum_action}): Consider the {\it classical} equation of
motion (\ref{classical_equation}) with the $\delta$-correlated
noise $f_{\rm th}({\bf z}, t)$; we search for the quantum
Hamiltonian that reproduces the same equation of motion in the
classical limit. One class of possible Hamiltonians has been
introduced by Caldeira and Leggett \cite{Caldeira} where the
system under consideration (an elastic manifold in our case) is
assumed to be coupled to a bath consisting of harmonic
oscillators. The Hamiltonian of the bath can be written in the
form
\begin{equation}
   {\hat H}_{\rm bath}=\int\thinspace d^{d}z
   \sum_{j}\biggl[\frac{{\hat P}_{j}^{2}({\bf z})}{2M_{j}({\bf z})}
   +\frac{M_{j}({\bf z})}{2}\Omega_{j}^{2}({\bf z})
   {\biggl( X_{j}({\bf z})-\frac{c_{j}({\bf z})}
   {M_{j}({\bf z})\Omega_{j}^{2}({\bf z})}
   {u({\bf z})}\biggr)}^2\biggr],
\end{equation}
where each segment of the manifold ${\bf z}$ is coupled to the
system of harmonic oscillators. This Hamiltonian contributes to
the quantum-mechanical action with
\begin{eqnarray}
   S_{\rm bath}[{\bf X}] + S_{\rm int} [u, {\bf X}]
   &=& \int\thinspace dt\int\thinspace d^{d}z \sum_{j}
   \biggl[\frac{M_{j}({\bf z})}{2}{{\dot X}_{j}^{2}({\bf z})}
   \nonumber \\
   &-&\frac{M_{j}({\bf z})}{2}\Omega_{j}^{2}({\bf z})
   {\biggl( X_{j}({\bf z})-\frac{c_{j}({\bf z})}
   {M_{j}({\bf z})\Omega_{j}^{2}({\bf z})}{u({\bf z})}\biggr)}^2
   \biggr],
\end{eqnarray}
where ${\bf X}$ is the vector with components $X_j$ describing the
oscillators of the bath. The action corresponding to the elastic
manifold $S_{0}[u]$ has the form
\begin{equation}
   S_{0}[u] = - \int \thinspace dt\int \thinspace d^{d}z
   \biggl[\frac{c}{2}
   {\left(\frac{\partial u}{\partial{\bf z}}\right)}^2
   -\frac{\rho}{2}{\left(\frac{\partial u}{\partial t}\right)}^2
   + U(u,{\bf z})-fu\biggr].
\end{equation}
After substituting the action $S =  S_{\rm bath} + S_{\rm int}
+S_{0}$ into Eq.~(\ref{quantum_partition_function}) we can
eliminate the bath degrees of freedom appearing only quadratically
in the action. We define the spectral density
\begin{equation}
   J (\omega ) \equiv \frac{\pi}{2}
   \sum\limits_{j}\frac{c_{j}^{2}({\bf z})}{M_{j}({\bf z})
   \Omega_{j}({\bf z})}\delta(\omega-\Omega_j ), \quad \omega > 0;
\end{equation}
the ohmic kernel $J(\omega) = \eta \omega$ produces the action
(\ref{quantum_action}). In the classical limit (high temperatures)
one can {\it rigorously} expand the potential energy terms
in (\ref{quantum_action})
in
${\tilde y}$  and obtain the
action (\ref{MSR_action}) which is equivalent to
Eq.~(\ref{classical_equation}). Note that we have introduced new
coordinates ${\tilde u}$ and ${\tilde y}$ in the action
(\ref{quantum_action}) corresponding to the ``center of mass'' and
relative motion of the trajectories in
(\ref{quantum_partition_function}).

\section{Fixed-point Function of the Correlator}
\label{appendix1}

In this appendix we derive the relation (\ref{useful_relation}).
We investigate the static RG fixed-point in the absence of a
driving force; the temperature and quantum fluctuations
renormalize to zero, $C_l^{>}(\Lambda, t=0) \rightarrow 0$, and
hence we need to consider the function $\Delta_l(u)$ alone. For $l
< l_c$ we can neglect the nonlinear terms on the right-hand side
of (\ref{correlator_rg}). Solving the remaining linear equation we
obtain
\begin{equation}
   \Delta_{l_c}\sim \Delta (0) e^{(\epsilon - 2\zeta )l_c}
\end{equation}
for the correlator height and $\xi^\ast \approx \xi \exp(-\zeta
l_c)$ for its width (the latter result follows from integration of
the second term in (\ref{correlator_rg})). On the other hand, for
$l \gg l_c$ the correlator is close to its fixed-point value.
Using equation (\ref{correlator_rg}) for $\partial_l\Delta_l (u) =
0$ and substituting $u=0$ we obtain
\begin{equation}
   \left (\epsilon - 2\zeta \right )\Delta_{l_c} (0)
   \sim  \left (\epsilon - 2\zeta \right )\Delta^{\ast} (0)
   \sim I{{\Delta^\ast}^{\prime}}^2 (+0)
\end{equation}
and replacing $I = A_d \Lambda^d/c^2\Lambda^4$ and $\zeta \sim
\epsilon$ we obtain (\ref{useful_relation}). Combining these
results with $|{\Delta^\ast}^{\prime}(0+)| \sim \Delta^\ast
(0)/\xi^\ast$ we find the estimates $\Delta^\ast (0) \sim
(c\Lambda^2)^2\xi^2\Lambda^{-d}e^{-2\zeta l_c}$ and
$|{\Delta^\ast}^\prime(0+)| \sim
(c\Lambda^2)^2\xi\Lambda^{-d}e^{-\zeta l_c}$.

\section{Dangerous Relevant Variables}
\label{appendix2}

As mentioned in section~\ref{random_friction}, the RG flow
generates dangerous {\it relevant} variables corresponding to the
cumulants of the friction distribution\cite{RBF}. From a physical point of
view, the friction coefficient $\eta$ in the equation of motion
(\ref{classical_equation}) is always spatially inhomogeneous due
to the presence of disorder, so that the bare values of the
cumulants describing the friction distribution are always nonzero.
It turns out that under the RG transformation these cumulants grow
extremely rapidly and the description of the thermally activated
or quantum motion in terms of the renormalized viscosity $\eta_l$
or the function $D_{l}(\omega )$ (see section~\ref{full_analysis})
alone is not appropriate. Since the renormalized viscosity on a
certain lengthscale corresponds to the waiting time on this scale,
we conclude that the broad distribution of the friction (it is
broad as the cumulants grow very rapidly) implies a broad
distribution of waiting times. Hence, the proper description of
the problem requires renormalization of the waiting time
probability distribution function rather than its first moment
alone. The main goal of this appendix is to show how the dangerous
variables are renormalized under the RG flow. We will also show
that even if initially all of them are zero, they will be
generated by the random pinning. We will restrict ourselves to the
case of high temperatures; the same variables will be generated in
the quantum case.

Let us assume that the friction $\eta$ in the equation of motion
(\ref{classical_equation}) is a spatially inhomogeneous function
of the coordinate; in general, $\eta =\eta (u, {\bf z})$. For
simplicity, we neglect the dependence on $u$ (this does not change
the result qualitatively). A physical realization where this
approximation is valid is the pinned charge-density-wave: the
friction $\eta (u, {\bf z})$ is a periodic function of $u$ and,
hence, can be expanded in a Fourier series; $\eta ({\bf z})$
then is the $u$-independent (the zeroth) harmonic.
Below we will assume that the $\eta$-disorder and the
$\Delta$-disorder are uncorrelated and short-ranged; in reality,
these two types of disorder should be correlated.
If we take these correlations into account,
the dangerous variables will still exist and the RG flow
will be similar. In this section we assume that the response
function has the form $1/\left (c{\bf k}^2 -i\eta_l\omega\right )$,
i.e., we do not consider the full dynamic response associated with
$D_l (\omega )$; this approach is valid in the classical case,
see sections~\ref{loc_tr} and \ref{cl_cr}.

\begin{figure}
\centerline{\epsfxsize = 7.0cm \epsfbox{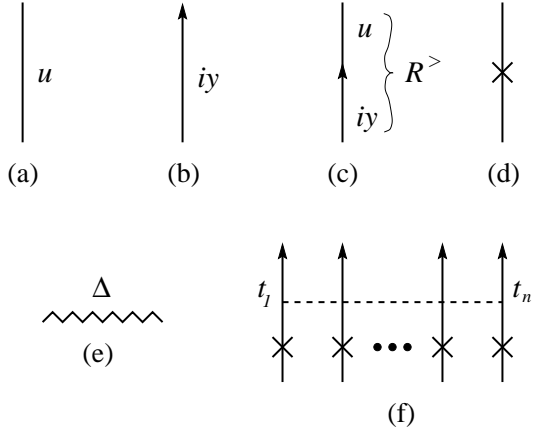}}
\vspace{0.3truecm} \caption{Elements used in diagrams: (a) the
displacement field $u$; (b) the $iy$-field; (c) the response
function $R^>$; (d) crosses denote derivatives, with respect to time
in the $u$-field if the cross is on a solid line, or with respect
to the time difference in the response function if the cross is on
a response function line; (e) diagrammatic representation of the
correlator $\Delta$; (f) diagrammatic representation of random
friction terms $(-1)^n \eta^{(n)} \prod {\dot u}({\bf z}, t_k)
iy({\bf z}, t_k)$, see (\ref{A_rand}); dots $\dots$ stand for
further omitted lines.} \label{dictionary}
\end{figure}

\begin{figure}
\centerline{\epsfxsize = 6.0cm \epsfbox{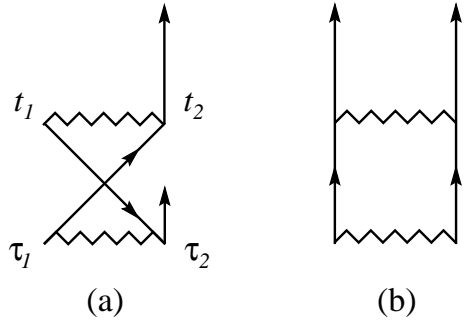}}
\vspace{0.3truecm} \caption{Diagrams describing the generation
of the second cumulant of the friction distribution;
their low-frequency expansions feed back to the
second cumulant.} \label{second_cumulant_diagrams}
\end{figure}

Averaging over the disorder in $\eta$ within the MSR-functional
(\ref{MSR_action}) we see that the friction term will be
transformed into
\begin{eqnarray}
   &-&\int d^{d}{z}dt\thinspace\thinspace \eta
   {\dot u}({\bf z},t) iy({\bf z},t)
   \rightarrow -\int d^{d}{z}dt\thinspace\thinspace \eta
   {\dot u}({\bf z},t) iy({\bf z},t)
   + A_{\rm rand}\nonumber\\
   &\equiv&
   -\int\thinspace d^{d}{ z}dt\thinspace\thinspace \eta
   {\dot u}({\bf z}, t) iy ({\bf z}, t)
   + \sum_{k\ge 2}A^{(k)}\label{A_rand}\\
   &=&-\int d^{d}{z}dt \thinspace\thinspace \eta
   {\dot u}({\bf z}, t) iy ({\bf z}, t)
   +\sum_{k\ge 2} (-1)^{k}\eta^{(k)}\int d^{d}z \prod
   \limits_{i=1}^{k}dt_{i}{\dot u}({\bf z},t_i)iy({\bf z},t_i).
   \nonumber
\end{eqnarray}
When deriving the RG equations we consider the term $A_{\rm rand}$
and the usual disorder-induced term
\begin{equation}
   A_{\rm dis}= \frac{1}{2} \int d^d z d\tau_1 d\tau_2 \thinspace
   \Delta\left[u({\bf z}, \tau_1) - u({\bf z}, \tau_2) \right]
\end{equation}
as perturbations. To one-loop order we have to calculate the
second cumulant of the perturbation. There will be a cross-term of
the form $\delta\langle A_{\rm rand} A_{\rm dis}\rangle_{>}$,
where $\delta\langle H \rangle_{>} $ denotes the change in $H$
after averaging over fast modes. Let us show that the average of
this form gives rise to the singular renormalization of the
coefficients $\eta^{(n)}$. Figure \ref{dictionary} summarizes the
various symbols appearing in the diagrammatic expansion: Lines
without arrows denote displacement fields $u$, lines with arrows
at the end correspond to the auxiliary fields $iy$, and lines with
arrows in the middle denote response functions. Crosses stand for
time derivatives and the wavy horizontal lines denote disorder
correlators $\Delta_l\left[u({\bf z},t_1)-u({\bf z},t_2)\right]$.
Dashed horizontal lines stand for $(-1)^k \eta^{(k)}$ and dots
$\dots$ represent further omitted lines. For each response
function with an arrow coming into a wavy vertex (see e.g., the
diagrams in Figs.~\ref{second_cumulant_diagrams}(a) and (b)), the
disorder correlator $\Delta[u({\bf z}, t_1)-u({\bf z},t_2)]$ is
differentiated with respect to its argument. These derivatives
appear as a consequence of averaging of the correlator and
the $iy$-fields, see Eq.~(\ref{useful_formula_2}) below.
Dashed and wavy lines in diagrams connect times, the upper
line connects $t_1$, $t_2$, $\dots$, and the lower line connects
the times $\tau_1$ and $\tau_2$. Note that there is no time
ordering corresponding to vertical or horizontal directions. The
fields at the upper vertices in diagrams are taken at the spacial
coordinate ${\bf z}$ and the lower vertex is at ${\bf
z}^{\prime}$. The argument of the response function involves the
differences between the times and spatial coordinates of the end
and starting points. We quote three useful formulae
\begin{equation}
   \delta\langle u({\bf z},t)iy({\bf z}^{\prime},t^{\prime})\rangle_>
   = R_l^{>} ({\bf z} - {\bf z}^{\prime}, t- t^{\prime}) dl
\label{useful_formula_1}
\end{equation}
\begin{equation}
   \delta\left\langle
   \Delta \left[u({\bf z},\tau_1)-u({\bf z},\tau_2)\right]
   iy ({\bf z}^{\prime}, t^{\prime}) \right\rangle_>
   =\Delta^{\prime}\left[u({\bf z},\tau_1)
   -u({\bf z},\tau_2)\right]
   [R_l^{>}({\bf z}-{\bf z}^{\prime},\tau_1-t^{\prime})
   -R_l^{>}({\bf z}-{\bf z}^{\prime},\tau_2-t^{\prime})]dl
\label{useful_formula_2}
\end{equation}
\begin{equation}
   \int d^dz^\prime R_l^{>}({\bf z}^\prime,t_1)
   R_l^{>}({\bf z}^\prime,t_2)
   =\frac{A_d\Lambda^d}{{(2\pi)}^d}\,dl\,
   R_l (\Lambda,t_1) R_l (\Lambda,t_2),
\label{useful_formula_3}
\end{equation}
where $R_l(\Lambda,t)$ denotes the (partly Fourier transformed)
response function (\ref{definition_r}) and $R^{>}_l (\Lambda , t)$
is defined in (\ref{definition_r_>}).

\begin{figure}
\centerline{\epsfxsize = 6.0cm \epsfbox{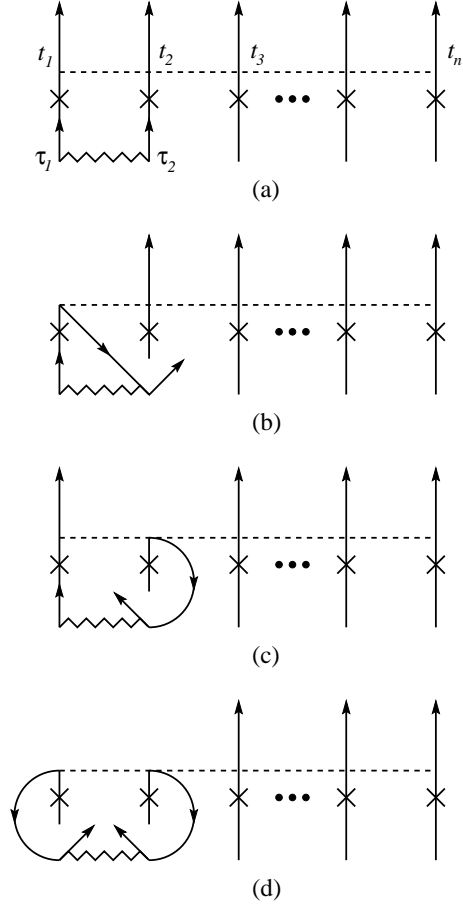}}
\vspace{0.3truecm} \caption{Dangerous diagrams contributing to the
renormalization of the cumulants of the friction-coefficient
distribution. These terms arise from averaging over fast modes in
$\langle A_{\rm rand} A_{\rm dis}\rangle_{>}$: (a) with
multiplicative factors $n(n-1)$ (due to the permutation symmetry),
(b) with multiplicative factor $2n$, (c) with multiplicative
factor $2n(n-1)$, (d) with multiplicative factor $n(n-1)/2$. For
convenience we draw some of the $iy$-fields and response functions
at a $45$ degrees angle or deform the lines.}
\label{dangerous_diagrams}
\end{figure}

The contribution $\delta D_{1}$ of the diagram in Fig.\
\ref{dangerous_diagrams}(a) has the form,
\begin{eqnarray}
   \delta D_{1} &=& \frac{1}{2}
   \eta_{l}^{(n)}\int d^{d}z d^dz^{\prime}
   \int d\tau_1 d\tau_2\thinspace
   \left[\prod\limits_{k=1}^{n}dt_k\right ]
   \delta\left\langle
   {\dot u}({\bf z}, t_1)iy({\bf z}^{\prime}, \tau_1)
   \right\rangle_{>}
   \delta\left\langle
   {\dot u}({\bf z}, t_2)iy({\bf z}^{\prime}, \tau_2)
   \right\rangle_{>}
   \nonumber\\
   &&\qquad\qquad \times iy({\bf z}, t_1 )
   iy({\bf z}, t_2 )
   \Delta_l\left[u\left ({\bf z}^{\prime} ,\tau_1 \right )
   -u\left ({\bf z}^{\prime}, \tau_2 \right ) \right]
    \prod\limits_{k=3}^{n}
   {\dot u}({\bf z}, t_k)
   iy({\bf z}, t_k)
   \label{1_diagram_random_p}\\
   &=&\frac{1}{2} \eta_l^{(n)}\frac{A_d\Lambda^d}{{\left (2\pi \right )}^{d}}dl
   \int d^d z \left [ \prod\limits_{k=1}^{n}dt_k \right ]
   {\dot R}_l(\Lambda , t_1 - \tau_1 )
   {\dot R}_l(\Lambda , t_2 - \tau_2 )
   iy({\bf z}, t_1 )
   iy({\bf z}, t_2 )
   \nonumber\\
   &&\qquad\qquad\times\int d\tau_1 d\tau_2\thinspace
   \Delta_l\left[u\left ({\bf z},\tau_1 \right )
   - u\left ({\bf z}, \tau_2 \right )\right]
   \prod\limits_{k=3}^{n}
   {\dot u}({\bf z}, t_k)
   iy({\bf z}, t_k).\nonumber
\end{eqnarray}
In the last equation, we have used (\ref{useful_formula_1}) in
order to express averages over fast modes through response
functions and have integrated over ${\bf z}^{\prime}$ using
(\ref{useful_formula_3}). We rewrite the product of the two
response functions as ${\dot R}_l(\Lambda,t_1-\tau_1) {\dot R}_l
(\Lambda,t_2-\tau_2)=\partial_{\tau_1}{R}_l(\Lambda,t_1-\tau_1)
\partial_{\tau_2} {R}_l(\Lambda,t_2-\tau_2)$ and integrate by
parts over variables $\tau_1$ and $\tau_2$, thus generating the
second derivative of the correlator $\Delta$. The response
functions connect times $t_1$ and $\tau_1$ and $t_2$ and $\tau_2$.
Consequently, we can set $t_1\approx \tau_1$ and $t_2\approx
\tau_2$ in the time arguments of the $iy$-fields and integrate
over the variables $t_1$ and $t_2$ in the response functions using
the formula $\int\thinspace dt R (\Lambda , t) = R(\Lambda ,
\omega = 0) = 1/c\Lambda^2$. In the end we obtain an expression
involving the integrals $\int d^dz \thinspace d\tau_1 \thinspace d
\tau_2 \thinspace dt_3\dots dt_n$ and performing substitution
$\tau_1 \rightarrow t_1$ and $\tau_2 \rightarrow t_2$ we arrive at
the final contribution of the diagram~\ref{dangerous_diagrams}(a)
\begin{eqnarray}
   \delta D_{1} = - \frac{1}{2}
   \eta_l^{(n)}\frac{A_d\Lambda^d}{{\left (2\pi \right )}^{d}}
   \frac{dl}{{\left(c\Lambda^2\right )}^{2}} \int d^d z \left [
   \prod\limits_{k=1}^{n}dt_k {\dot u}({\bf z}, t_k) iy({\bf z}, t_k)
   \right ] \Delta_{l}^{\prime\prime}\left[u\left ({\bf z},t_1
   \right ) - u\left ({\bf z}, t_2 \right ) \right].
   \label{1_diagram_random}
\end{eqnarray}
Note that in the diagrammatic language integrating by parts
corresponds to drawing $u$-field lines out of the wavy vertex and
moving crosses (i.e., time-derivatives) from the response
functions down to the displacement fields. The low-frequency
expansion of the diagram in Fig.~\ref{dangerous_diagrams}(a) then
takes the form of the diagram in Fig.~\ref{dictionary}(f) [this
low-frequency analysis involves the expansion of the argument
$u({\bf z}, \tau_1 ) - u({\bf z}, \tau_2 )$ of the correlator
$\Delta$ in a Taylor series in powers of $(\tau_1 - \tau_2)$; the
first term (equal to zero) of this expansion produces the desired
contribution]. There are $n(n-1)$ equivalent diagrams of the type
in Fig.\ \ref{dangerous_diagrams}(a) allowed by the permutation
symmetry.

The expression corresponding to the diagram in
Fig.~\ref{dangerous_diagrams}(b) is given by
\begin{eqnarray}
   \delta D_{2}&=& \frac{1}{2}\eta_l^{(n)}
   \int d^{d}z d^{d}z^{\prime} \int d\tau_1 d\tau_2\thinspace
   \left [\prod\limits_{k=1}^{n}dt_k\right ]
   \delta\langle {\dot u} ({\bf z}, t_1)iy ({\bf z}^{\prime},\tau_1 )
   \rangle_>\delta\langle
   \Delta_l\left[u\left ({\bf z}^{\prime} ,\tau_1 \right)
   -u\left ({\bf z}^{\prime}, \tau_2 \right)\right]
   iy ({\bf z}, t_1 )\rangle_>\nonumber\\
   &&\qquad\qquad\times iy({\bf z}^{\prime},\tau_2)
   \prod\limits_{k=2}^{n}
   {\dot u}({\bf z}, t_k )iy ({\bf z}, t_k )\label{2_diagram_random}\\
   &=&-\frac{1}{2}
   \eta_l^{(n)}\Delta_{l}^{\prime\prime}(0)
   \frac{A_d\Lambda^d}{{\left (2\pi \right )}^{d}}
   \frac{dl}{{\left (c\Lambda^2\right )}^{2}}
   \int d^d z
   \prod\limits_{k=1}^{n} dt_k{\dot u}({\bf z}, t_k)
   iy({\bf z}, t_k);\nonumber
\end{eqnarray}
there are $2n$ topologically equivalent diagrams of this class.
Note that in contrast to the term $\delta D_1$ the argument of the
function $\Delta_l (u)$ in the above expression is zero.
The diagram shown in Fig.~\ref{dangerous_diagrams}(c) gives the
contribution
\begin{eqnarray}
   \delta D_3 &=&
   \frac{1}{2}
   \eta_l^{(n)} \int d^{d}z  d^{d}z^{\prime}
   \int d\tau_1 d\tau_2\thinspace
   \left[\prod\limits_{k=1}^{n}dt_k \right]
   \delta\langle {\dot u} ({\bf z}, t_1)iy ({\bf z}^{\prime},\tau_1)
   \rangle_> \delta\langle \Delta\left[u\left ({\bf z}^{\prime},
   \tau_1 \right ) - u\left ({\bf z}^{\prime}, \tau_2 \right )
   \right] iy ({\bf z}, t_2 )\rangle_>\nonumber\\
   &&\qquad\qquad\times iy ({\bf z},t_1){\dot u}({\bf z},t_2 )
   iy({\bf z}^{\prime},\tau_2)
   \prod\limits_{k=3}^{n}{\dot u}({\bf z}, t_k)
   iy({\bf z}, t_k) \label{3_diagram_random}\\
   &=& - \frac{1}{2}
    \eta_l^{(n)}\frac{A_d\Lambda^d}{{\left(2\pi \right)}^{d}}
   \frac{dl}{{\left (c\Lambda^2\right )}^{2}}\int d^d z
   \left[\prod\limits_{k=1}^{n}dt_k {\dot u}({\bf z}, t_k)
   iy({\bf z}, t_k) \right ] \Delta_{l}^{\prime\prime}
   \left[u\left ({\bf z},t_1\right)-u\left({\bf z},t_2\right)
   \right];\nonumber
\end{eqnarray}
there are $2n(n-1)$ topologically equivalent diagrams of this
class.
Finally, the diagram in Fig.~\ref{dangerous_diagrams}(d) is given
by the expression
\begin{eqnarray}
\delta D_4 & = &
   \frac{1}{2}\eta_l^{(n)}
   \int d^{d}z d^{d}z^{\prime} \int d\tau_1 d\tau_2\thinspace
   \left [\prod\limits_{k=1}^{n}dt_k\right ]
   {\dot u}({\bf z}, t_1) {\dot u}({\bf z}, t_2)
   iy({\bf z}^{\prime}, \tau_1)
   iy({\bf z}^{\prime}, \tau_2)
   \nonumber\\
   &&\quad\quad
   \times \left [ \prod_{k=3}^{n}
   {\dot u}({\bf z}, t_k) iy({\bf z}, t_{k}) \right ]
   \delta\left\langle iy({\bf z}, t_1)
   \delta\left\langle
   \Delta\left[u\left ({\bf z}^{\prime},
   \tau_1 \right ) - u\left ({\bf z}^{\prime}, \tau_2 \right )
   \right] iy({\bf z}, t_2)
   \right\rangle_>
   \right\rangle_{>}
   \nonumber\\
   & = &
   \frac{1}{2}\eta^{(n)}_l
    \frac{A_d\Lambda^d}{{\left (2\pi \right )}^{d}}
   \frac{dl}{{\left (c\Lambda^2\right )}^2}
   \int d^d z \left [ \prod\limits_{k=1}^{n}dt_k \right ]
   d\tau_1 d\tau_2
   {\dot u}({\bf z}, t_1) {\dot u}({\bf z}, t_2)
   iy({\bf z}, \tau_1) iy({\bf z}, \tau_2)
   \left [\prod_{k=3}^{n}
   {\dot u}({\bf z}, t_k) iy({\bf z}, t_{k})
   \right ]
   \nonumber\\
   & \times &
   \{R(\Lambda , \tau_1 - t_1)R(\Lambda , \tau_1 - t_2)
   + R(\Lambda , \tau_2 - t_1)R(\Lambda , \tau_2 - t_2)
   \nonumber\\
   &&
   - R(\Lambda , \tau_1 - t_1)R(\Lambda , \tau_2 - t_2)
   - R(\Lambda , \tau_1 - t_2)R(\Lambda , \tau_2 - t_1) \}
   \Delta^{\prime\prime}\left[u\left ({\bf z},
   \tau_1 \right ) - u\left ({\bf z}, \tau_2 \right ) \right]
   = 2 \delta D_3 .
   \label{4_diagram_random}
\end{eqnarray}
The first two terms in the curly brackets do not feed back to
the random friction cumulants, while the third and fourth terms give
the desired contribution. The multiplication factor of the diagram
in Fig.~\ref{dangerous_diagrams}(d) is equal to $n(n-1 )/2$.

In order to find the contribution to $\eta^{(n)}_{l}$ we
need to set the argument $u({\bf z},t_1)-u({\bf z},t_2)$ in the
correlator $\Delta (u)$ to zero in (\ref{1_diagram_random}),
(\ref{3_diagram_random}), and  (\ref{4_diagram_random})
(as only this term feeds back to the random friction cumulants)
and sum over the four contributions (\ref{1_diagram_random}),
(\ref{2_diagram_random}), (\ref{3_diagram_random}), and
(\ref{4_diagram_random}) each multiplied with its appropriate
weight; the result takes the form
\begin{equation}
   \delta\eta_l^{(n)} = - \eta_l^{(n)}\left (2n^2-n\right)
   \frac{A_d\Lambda^d}{{\left (2\pi \right )}^{d}}
   \frac{\Delta_l^{\prime\prime}(0)}
   {{\left (c\Lambda^2\right )}^2}dl.
\end{equation}
Next, we show that the term $A_{\rm dis}$ itself generates
the second cumulant $\eta^{(2)}$. The dangerous
relevant diagrams arise from the average $\langle A_{\rm dis}
A_{\rm dis}\rangle_{>} $, see Fig.~\ref{second_cumulant_diagrams};
their low-frequency expansion gives the contribution to the random
friction. The contribution of the diagram
\ref{second_cumulant_diagrams}(a) takes the form
\begin{eqnarray}
   \delta {\tilde D}_1 &=&
   \frac{1}{8}\frac{A_d\Lambda^d}{{\left (2\pi \right )}^{4}}dl
   \int d^d z dt_1 dt_2\thinspace\thinspace
   \Delta_l^{\prime} \left[u({\bf z},t_1)-u({\bf z},t_2)\right]
   iy ({\bf z}, t_2 ) iy ({\bf z}, \tau_2 )\nonumber\\
   &&\qquad\qquad \times \int d\tau_1 d\tau_2 \thinspace
   \Delta_l^{\prime} \left[u({\bf z}, \tau_1 )
   - u ({\bf z}, \tau_2 )\right]
   R_l(\Lambda, t_2 - \tau_1 )
   R_l(\Lambda, \tau_2 - t_1 ),
   \label{second_cumulant_diagram_1}
\end{eqnarray}
where the averaging over fast modes and the integration over ${\bf
z}^{\prime}$ has already been carried out. The response functions
connect the times $t_2,~\tau_1$ and $\tau_2,~t_1$ and we can
expand
\begin{eqnarray}
   u({\bf z}, t_1 )- u ({\bf z}, t_2 ) & = &
   u({\bf z}, \tau_2 + (t_1 - \tau_2)) -
   u({\bf z}, t_2)
   \nonumber\\
   & \approx  & u({\bf z}, \tau_2) - u({\bf z}, t_2) +
   {\dot u}({\bf z}, \tau_2 ) (t_1 - \tau_2 );
\end{eqnarray}
similarly, $u({\bf z}, \tau_1 )- u ({\bf z}, \tau_2 )\approx
u({\bf z}, t_2 )- u ({\bf z}, \tau_2 ) + {\dot u} ({\bf z}, t_2 )
(\tau_1 - t_2 )$. Substituting these expansions into
(\ref{second_cumulant_diagram_1}) and expanding the correlators
$\Delta^{\prime} (u)$ we arrive at the expression (note that $\int
dt \thinspace t R(\Lambda , t) = \eta_l/{\left (c\Lambda^2\right
)}^2 $)
\begin{eqnarray}
   \delta {\tilde D}_1 &=&
   \frac{1}{8}\frac{A_d\Lambda^d}{{\left (2\pi \right )}^{d}}
   \frac{\eta_l^2}{{\left (c\Lambda^2 \right )}^4}dl
   \nonumber\\
   &&\qquad\times \int  d^d z dt_1 dt_2\thinspace
   {\Delta_l^{\prime\prime}}^2 \left[u({\bf z}, t_1 )
   -u({\bf z}, t_2 )\right]
   {\dot u}({\bf z}, t_1 )
   iy ({\bf z}, t_1 )
   {\dot u}({\bf z}, t_2 )
   iy ({\bf z}, t_2 );
\end{eqnarray}
there are 4 topologically equivalent diagrams of this class.

Diagram~(b) in Fig.~\ref{second_cumulant_diagrams} gives the
contribution
\begin{eqnarray}
   \delta {\tilde D}_2 &=&
   \frac{1}{8}\frac{A_d\Lambda^d}{{\left (2\pi \right )}^{d}}
   dl\int d^dz dt_1 dt_2
   \thinspace
   \Delta_l^{\prime\prime} \left[u({\bf z}, t_1 ) -
   u({\bf z}, t_1 )\right] iy ({\bf z}, t_1 )
   iy ({\bf z}, t_2 )\nonumber\\
   &&\qquad\times \int d\tau_1 d\tau_2\thinspace
   \Delta_l \left[u({\bf z}, \tau_1 ) -
   u({\bf z}, \tau_2 )\right] R_l(\Lambda , \tau_1 - t_1)
   R_l(\Lambda , \tau_2 - t_2),
   \label{second_cumulant_diagram_2}
\end{eqnarray}
where the averaging over fast modes and the integration over ${\bf
z}^{\prime}$ again has been performed already. The response
functions connect the points $\tau_1$ and $t_1$ and $\tau_2$ and
$t_2$. Expanding the displacement fields in $t_1 - \tau_1$ and
$\tau_2 - t_2$, substituting them into
(\ref{second_cumulant_diagram_2}), and expanding the correlator
$\Delta (u)$ we see that $\delta {\tilde D}_2$ = $\delta {\tilde
D}_1$. There are 4 topologically equivalent diagrams of class (b).
Summing up the contributions $\delta {\tilde D}_1$ and $\delta
{\tilde D}_2$ and multiplying the sum by 4 we obtain the
contribution $\delta{\tilde \eta}^{(2)}_l$ of the disorder term
to the second cumulant,
\begin{equation}
   \delta {\tilde \eta}^{(2)}_l =
   \frac{A_d\Lambda^d}{{\left (2\pi \right )}^{d}}
   \frac{\eta_l^2}{{\left (c\Lambda^2 \right )}^4}
   {\Delta^{\prime\prime}_l}^2 (0) dl.
   \label{contribution_second_cumulant}
\end{equation}

Finally, we can write the one-loop RG equation for the cumulants
of the random friction distribution; accounting for the terms
(\ref{1_diagram_random}), (\ref{2_diagram_random}),
(\ref{3_diagram_random}),
(\ref{4_diagram_random}),
and (\ref{contribution_second_cumulant})
we find the flow equations
\begin{eqnarray}
   \partial_l \eta_l^{(2)}&=&
   \left(4-d-2z \right )\eta_l^{(2)} + I\frac{\eta_{l}^2}
   {{\left({c\Lambda^2}\right )}^4 }{\Delta_l^{\prime\prime}}^2 (0)
   -\frac{6I}{ {\left (c\Lambda^2 \right )}^2}
   \Delta^{\prime\prime}_{l}(0)\eta_l^{(2)},
   \label{equation_random_2}\\
   \partial_l\eta_l^{(n)} &=&
   \left(d+2n-dn-zn\right)\eta_l^{(n)}
   -\left (2n^2 - n \right )
   \frac{I}{ {\left (c\Lambda^2 \right )}^4}
   \Delta^{\prime\prime}_{l}(0)\eta_l^{(n)}, \qquad n \ne 2.
   \label{equation_random_3}
\end{eqnarray}
The system of RG equations
(\ref{equation_random_2}) and (\ref{equation_random_3}) has been
derived to lowest order in $4 - \epsilon$. To next order, the
second cumulant will generate the third cumulant. To third order,
the third cumulant will generate the fourth cumulant, etc., i.e.,
{\it all} cumulants will be generated in the RG flow even if all
of them (except for the first one) are equal to zero initially.
From a physical point of view, the friction will always be random
as the point-like impurities suppress the order parameter randomly
and, hence, the dynamic characteristics of the medium (e.g., a
superconductor) is random as well. Since $-\Delta_l^{\prime\prime}
(0)\propto {1}/{T_l}\propto e^{\theta l}$ and the correction due
to disorder grows as $n^2$ for large $n$, the random friction
probability distribution becomes very broad and one needs to take
into account {\it all} its moments and not just the friction
$\eta_l$ only.  In order to obtain Eqs.~(\ref{equation_random_2})
and (\ref{equation_random_3}) we have
made the approximation $u({\bf z}, t_1 ) - u ({\bf z}, t_2 ) = 0$
in the final expressions for $\delta D_1$, $\delta D_3$, $\delta
{\tilde D}_1$, and $\delta {\tilde D}_2$;
in general, however, one needs to renormalize the full
functional (\ref{super_dangerous_new}). During the RG procedure,
other terms will be generated and it is unclear at this stage how
to take all these terms into account in a controllable way.

\end{appendix}


\begin{thebibliography}{99}
\bibitem{colored} By colored noise we mean any stochastic force
whose Fourier transform of the correlator is not a constant
in the $\omega$-space.
\bibitem{v-scales} Another open question is related to the solution
of the RG equations in section~\ref{solution_of_rg_equations}.
>From the mathematical point of view it is not easy to describe the
crossover from the regime $u_{\rm vel}\ll u_{\rm smear}$ to
$u_{\rm vel}\gg u_{\rm smear}$. We expect that the behavior of the
manifold on large scales $L\gg L_f$ will be nontrivial.
In particular for the random bond disorder, the wandering exponent
on scales $L\alt L_f$ and $L \agt L_f$ will be different.
This, however, will not affect the results for the velocity.
\end{thebibliography}
\end{document}